%% file: ms.tex
\documentclass[apj]{emulateapj}
\usepackage{graphics,longtable,rotating,lscape,enumerate, amsmath}

\newcommand{\oiiw}{\mbox{[\ion{O}{2}] $\lambda$3727}}
\newcommand{\niw}{\mbox{[\ion{N}{1}] $\lambda$5197,5200}}

\newcommand{\niibw}{\mbox{[\ion{N}{2}] $\lambda$6583}}

\newcommand{\oiiibw}{\mbox{[\ion{O}{3}] $\lambda$5007}}

\newcommand{\mgii}{\mbox{\ion{Mg}{2}}}
\newcommand{\oi}{\mbox{[\ion{O}{1}]}}
\newcommand{\oii}{\mbox{[\ion{O}{2}]}}
\newcommand{\nii}{\mbox{[\ion{N}{2}]}}
\newcommand{\hal}{\mbox{H$\alpha$}}

\newcommand{\hb}{\text{H$\beta$}}
\newcommand{\oiii}{\text{[\ion{O}{3}]}}

\newcommand{\Msun}{\hbox{$\rm M_\odot$}}
\newcommand{\hii}{\hbox{\ion{H}{2}}}
\newcommand{\hard}{\hbox{2--10~keV}}
\newcommand{\soft}{\hbox{0.5--2~keV}}

\shorttitle{X-ray vs. optical selection of AGNs}
\shortauthors{Yan et al.}

\begin{document}
\title{AEGIS: Demographics of X-ray and Optically Selected AGNs}
\author{Renbin Yan$^{1}$, Luis C. Ho$^{2}$, Jeffrey A. Newman$^3$, Alison L. Coil$^{4,5}$, Christopher N. A. Willmer$^6$, Elise S. Laird$^7$, Antonis Georgakakis$^8$, James Aird$^4$, Pauline Barmby$^9$, Kevin Bundy$^{10,11}$, Michael C. Cooper$^{11,12}$, Marc Davis$^{10,13}$, S. M. Faber$^{14}$, Taotao Fang$^{15}$, Roger L. Griffith$^{16}$, Anton M. Koekemoer$^{17}$, David C. Koo$^{14}$, Kirpal Nandra$^7$, Shinae Q. Park$^{18}$, Vicki L. Sarajedini$^{19}$, Benjamin J. Weiner$^6$, S. P. Willner$^{18}$}

\affil{$^1$ Department of Astronomy and Astrophysics, University of Toronto, 50 St. George Street, Toronto, ON M5S 3H4, Canada; yan@astro.utoronto.ca}
\affil{$^2$ The Observatories of the Carnegie Institution for Science, 813 Santa Barbara Street, Pasadena, CA 91101, USA}
\affil{$^3$ Department of Physics and Astronomy, University of Pittsburgh, 3941 O'Hara Street, Pittsburgh, PA 15260, USA}
\affil{$^4$ Department of Physics and Center for Astrophysics and Space Sciences, University of California, San Diego, 9500 Gilman Dr., La Jolla, CA 92093, USA}
\affil{$^5$ Alfred P. Sloan Foundation Fellow}
\affil{$^6$ Steward Observatory, University of Arizona, 933 North Cherry Avenue, Tucson, AZ 85721, USA}
\affil{$^7$ Astrophysics Group, Blackett Laboratory, Imperial College London, Prince Consort Rd, London SW7 2AZ, UK}
\affil{$^8$ National Observatory of Athens, V. Paulou and I. Metaxa, 11532, Greece}
\affil{$^9$ Department of Physics and Astronomy, University of Western Ontario, 1151 Richmond St., London, ON N6A 3K7, Canada}
\affil{$^10$ Department of Astronomy, University of California, Berkeley, 601 Campbell Hall, Berkeley, CA 94720, USA}
\affil{$^{11}$ Hubble Fellow}
\affil{$^{12}$ Center for Galaxy Evolution, Department of Physics and Astronomy, University of California, Irvine, CA 92697, USA}
\affil{$^{13}$ Department of Physics, University of California, Berkeley, CA 94720, USA}
\affil{$^{14}$ UCO/Lick Observatory, Department of Astronomy and Astrophysics, University of California, Santa Cruz, CA 95064, USA}
\affil{$^{15}$ Department of Physics and Astronomy, University of California, Irvine, CA 92697}
\affil{$^{16}$ Jet Propulsion Laboratory, California Institute of Technology, MS 169-327, 4800 Oak Grove Dr., Pasadena, CA 91109, USA}
\affil{$^{17}$ Space Telescope Science Institute, 3700 San Martin Drive, Baltimore, MD 21218, USA}
\affil{$^{18}$ Harvard–Smithsonian Center for Astrophysics, 60 Garden Street, Cambridge, MA 02138, USA} 
\affil{$^{19}$ Department of Astronomy, University of Florida, Gainesville, FL 32611, USA}

\begin{abstract}
\noindent 
We develop a new diagnostic method to classify galaxies into 
AGN hosts, star-forming galaxies, and absorption-dominated galaxies by 
combining the \oiii/\hb\ ratio with rest-frame $U-B$ color. This can be used to 
robustly select AGNs in galaxy samples at intermediate redshifts ($z<1$). We 
compare the result of this optical AGN selection with X-ray selection 
using a sample of 3150 galaxies with $0.3<z<0.8$ and $I_{\rm AB}<22$,
selected from the DEEP2 Galaxy Redshift Survey and the All-wavelength 
Extended Groth Strip International Survey (AEGIS). Among the 146 X-ray sources
in this sample, 58\% are classified optically as emission-line AGNs, 
the rest as star-forming 
galaxies or absorption-dominated galaxies. The latter are also known as 
``X-ray bright, optically normal galaxies'' (XBONGs). Analysis of the 
relationship between 
optical emission lines and X-ray properties shows that the completeness of 
optical AGN selection suffers from dependence on the star formation rate 
and the quality of observed spectra. It also shows that XBONGs do not appear
to be a physically distinct population from other X-ray detected, 
emission-line AGNs. On the other hand, X-ray AGN 
selection also has strong bias. About 2/3 of all emission-line 
AGNs at $L_{\rm bol}> 10^{44}~{\rm erg s^{-1}}$ in our sample are not detected 
in our 200 ks {\it Chandra} images, most likely due to moderate or heavy absorption 
by gas near the AGN. The 2--7 keV detection rate of Seyfert 2s at 
$z\sim0.6$ suggests that their column density distribution and Compton-thick fraction are similar to that of local Seyferts. 
Multiple sample selection techniques are needed to obtain as 
complete a sample as possible.
\rightskip=0pt
\end{abstract}




\section{Introduction}\label{sec:intro}
Recently, it has been realized that nearly every massive galaxy bulge hosts a supermassive black hole (SMBH) whose mass is tightly correlated with the stellar velocity dispersion or the bulge mass \citep{Magorrian98,FerrareseM00,Gebhardt00}. The tightness of these correlations suggests that the growth of the SMBH is physically linked with the evolution of the host galaxy. When SMBHs grow by accretion, they will appear observationally as active galactic nuclei (AGNs). A complete census of AGNs, which includes both the rare high-luminosity quasars and the more typical low-luminosity AGNs, is essential for our understanding of SMBH-galaxy co-evolution. However, such a census is not yet available beyond the local universe, primarily due to three reasons. 


First, at higher redshifts, it is difficult to spatially isolate the nuclear regions of galaxies for AGN detection. Due to the smaller apparent sizes and the fainter flux levels of distant galaxies, spatially isolated nuclear spectroscopy studies such as that of \cite{HoFS95} are not feasible at high redshifts. Using the integrated light, the detectability of AGNs at high-$z$ unavoidably depends on host galaxy properties such as stellar mass \citep[e.g.][]{Moran02} and star formation rate (SFR). All AGN selection methods have such dependences, differing in the galaxy property involved and on the level of sensitivity. This issue has not been fully addressed in the literature.


Second, there is no single method that can select a complete sample of AGNs. In other words, no single method identifies all the AGNs found by other methods. Every method has its own bias. Besides the different dependences on host galaxy properties mentioned above, obscuration of the AGN light also causes different objects to be missed by different techniques. For example, the two methods commonly regarded as the most complete for AGN selection are optical emission-line selection and X-ray selection. Dust extinction throughout the host galaxy can dramatically reduce the observed optical emission-line luminosity and bias optical selection against edge-on disk galaxies. X-ray selection, while unaffected even by the worst levels of extinction in the host galaxy, is biased against sources in which the X-ray emission is heavily absorbed and/or Compton-scattered by dense gas clouds much closer to the central engine.

When optical AGN selection is compared with X-ray AGN selection, one type of inconsistency attracts special attention: objects generally referred to as ``X-ray bright, optically normal galaxies (XBONGs)'' or ``optically dull'' X-ray galaxies \citep{Elvis81, Fiore00, Mushotzky00, Barger01, Comastri02, Maiolino03, Brusa03, Szokoly04, Rigby06, Cocchia07, Civano07, Caccianiga07,Trump09}. These galaxies are bright in the X-ray, so bright that they are undoubtably AGNs. However, their optical spectra either show no emission lines at all or else emission lines having line ratios typical of star-forming galaxies. The nature of these sources has been hotly debated. Some have argued they could have an intrinsically weak narrow-line region due to large covering factor or radiatively inefficent accretion flow \citep{YuanN04}. Others have suggested that the missing emission lines are due to dilution by host galaxy light \citep{Moran02} or extinction in the host galaxy \citep{Rigby06}. We will investigate the nature of this population here. However, we distinguish those hosts showing star-forming-like emission-line spectra from those showing no emission lines, as the explanations are different. 

Lastly, the standard method used for spectroscopic identification of AGNs is currently observationally too expensive to use for large galaxy samples at $z>0.4$. In the local universe, the classification of AGNs and star-forming galaxies is usually achieved by the use of optical emission line ratio diagnostics \citep[e.g.,][]{BPT,VeilleuxO87}. The commonly-used diagram involves two sets of line ratios: \niibw/\hal\ and \oiiibw/\hb\ (Figure~\ref{fig:bpt_sloan}). However, at $z>0.4$, \nii\ and \hal\ are redshifted out of the optical window into the near infrared. Other available diagnostics involve either two emission lines separated by a large wavelength interval, such as \oiiw/\hb \citep{RolaTT97}, which is sensitive to extinction and also has a limited observable redshift range, or involve the relatively weak lines, such as \niw/\hb\ ratio, which limit its applicability. These factors have hindered the construction of a large, complete, narrow-line AGN sample beyond $z\sim0.4$.

This paper first establishes a new optical emission-line diagnostic method that avoids the use of \nii\ and \hal\ lines so that we can select emission-line AGNs at redshifts beyond $z\sim0.4$.  We then make use of the rich multi-waveband data sets enabled by the All-wavelength Extended Groth Strip International Survey (AEGIS) Collaboration and high-quality DEEP2 optical/near-IR spectra to compare the two major AGN selection methods at $0.3<z<0.8$: optical emission-line diagnostics and X-ray selection. In particular, we pay attention to objects that are inconsistently classified by the two methods. This paves the way to the construction of a more complete AGN sample.

As a by-product, a comparison between optical emission-line luminosities and X-ray luminosities of AGNs can also help us evaluate the absorbing column density distribution among AGNs. In particular, this helps to constrain the fraction of Compton-thick AGNs, which are required to explain the spectrum of the hard X-ray background \citep{Gilli07}. Many studies on local AGN samples \citep[e.g.,][]{Bassani99, Risaliti99} have shown that about 50\% of Type 2 AGNs are Compton-thick. However, the Compton-thick fraction at higher redshift is more uncertain and is hotly debated \citep{Daddi07,Fiore08, Donley08,Treister09, Georgantopoulos09, Georgakakis10, Park10}, partly due to the lack of an emission-line selected AGN sample beyond the local universe. Therefore, we hope to shed some light on this topic with our emission-line AGN sample.

Throughout the paper, we use a flat $\Lambda$CDM cosmology with $\Omega_m=0.3$. We adopt a Hubble constant of $H_0=70~{\rm km~s^{-1}~Mpc^{-1}}$ to compute luminosity distances. All magnitudes are expressed in the AB system. 






\section{Data}

\subsection{X-ray imaging and optical identification}
The AEGIS-X survey (\citealt{Laird09}, L09 hereafter) has obtained 200 ks exposures over the entire Extended Groth Strip (EGS; \citealt{Davis07}; see \S\ref{sec:deep2}) using {\it Chandra} ACIS-I. It covers an area of 0.67 ${\rm deg}^2$ and reaches a limiting flux of $5.3\times10^{-17} {\rm erg~cm^{-2}~s^{-1}}$ in the \soft\ band and $3.8\times10^{-16} {\rm erg~cm^{-2}~s^{-1}}$ in the \hard\ band at the deepest point. It provides a unique combination of depth and area, bridging the gap between the ultradeep pencil-beam surveys, such as the {\it Chandra} Deep Fields (CDFs) and the shallower, very large-area surveys. Its areal coverage is nearly three times larger than the CDF-North and South combined. Compared to the {\it Chandra} imaging in the Cosmic Evolution Survey field ({\it Chandra}-COSMOS; \citealt{Elvis09}), our survey is slightly deeper but covers a slightly smaller area. A detailed comparison of area and flux limits among these state-of-the-art X-ray surveys is given by \cite{Elvis09}. 

The data reduction and point source catalogs of AEGIS-X are presented by L09. The reduction basically followed the techniques described by \cite{Nandra05} with new calculations of the point spread function (PSF; see L09). L09 used a wavelet detection algorithm run with a low threshold of $10^{-4}$ to identify candidate X-ray detections. Counts and background estimates were then extracted within an aperture corresponding to the 70\% encircled energy fraction (EEF) for each candidate source and used to calculate the probability that the source is a spurious detection. All sources with a false-positive probability ($p$) less than $4\times10^{-6}$ in any of the bands (soft, hard, ultrahard, or full band) are included in the final catalog.
The X-ray count rate was estimated using the 90\% EEF aperture. Unlike L09, we converted the count rate to flux using a photon index of $\Gamma=1.9$, which is more appropriate for unabsorbed sources \citep[e.g.,][]{Nandra94,Nandra97}. We also assumed this power-law spectrum to measure $K$-corrections from the observed X-ray flux to the rest-frame bands. Both fluxes and luminosities in this paper are reported for the rest-frame bands: 0.5--2 keV for the soft band and 2--10 keV for the hard band. The hardness ratio is defined as ${\rm HR}=(H-S)/(H+S)$, where $S$ and $H$ are the observed counts in the 0.5--2 keV and 2--7 keV bands, respectively. The HRs were computed using a Bayesian method following \cite{Park06} and using the BEHR package (version 07-11-2008). They are not $K$-corrected.

For sources detected ($p < 4\times10^{-6}$) in some bands but not others, if the false-positive probability ($p$) in an undetected band is less than 0.01, we still make a flux measurement for that band. Throughout this paper, unless otherwise noted, detection refers to having $p<4\times10^{-6}$. In certain cases, we treat sources detected in other bands but having $4\times10^{-6}<p<0.01$ in the 2--7~keV band as ``detections'' to increase the size of the sample with \hard\ flux measurement. 

The AEGIS X-ray source catalog presented by L09 contains 1325 sources. All but 8 are inside the boundaries of the DEEP2 CFH12K photometry catalog \citep{CoilNK04}. Based on a maximum-likelihood matching method \citep{SutherlandS92}, 895 sources are uniquely matched\footnote{Two X-ray sources in the catalog are matched to the same DEEP2 object, both with ${\rm LR}>0.5$. These two are not considered as valid matches and are removed; visual inspection of both X-ray and optical images suggests that both X-ray sources are in fact components of a single extended X-ray-emitting halo around a compact group of galaxies.} to the DEEP2 photometric catalog with a Likelihood Ratio (LR) greater than 0.5 (L09), corresponding to a 68.0\% optical identification rate. The estimated contamination rate is $\sim 6\%$. If limited to $R_{\rm AB}<24.1$, which is the DEEP2 survey limit, the optical identification rate is 53.5\%.

\subsection{Optical spectroscopy}

\subsubsection{DEEP2}\label{sec:deep2}

DEEP2 is a galaxy redshift survey using the DEIMOS spectrograph on the Keck-II telescope (\citealt{Davis03}; J. A. Newman et al. in prep). It covered four widely separated fields totaling 3 ${\rm deg}^2$ on the sky down to a limiting magnitude of $R_{\rm AB} = 24.1$. 
In the EGS, the survey obtained $\sim17,775$ spectra with $12,651$ yielding reliable redshifts.
DEEP2 spectra typically cover approximately $6500--9200$ \AA\ with a resolution of $R\sim5000$. The high resolution enables good subtraction of atmospheric emission lines and thus yields better sensitivity for line detection in the target spectra. DEEP2 employed a slit width of $1\arcsec$, which corresponds to a physical transverse scale of 7.1 kpc at $z=0.7$. 

Out of the 895 optically identified X-ray sources, 
375 were observed as part of the DEEP2 survey, yielding 249 successful redshifts (66.4\%) including 5 stars and 244 galaxies. 

\subsubsection{MMT/Hectospec follow-up}
Because the sampling fraction of the DEEP2 Galaxy Redshift Survey is $\sim50\%$ in the EGS field, and the survey began before the X-ray observations were taken, not all X-ray sources with optical counterparts were targeted for spectroscopy. We therefore obtained additional spectra using the Hectospec fiber spectrograph on MMT for as many X-ray sources with optical counterparts as possible. The observations and data reduction are described in detail by \cite{Coil09}. The spectra have a resolution of 6 \AA\  and cover a wavelength range of approximately $4500--9000$ \AA. In total, we targeted optical counterparts for 498 X-ray sources with 288 yielding reliable redshifts, including 23 stars and 265 galaxies. The redshift success rate is a strong function of the optical magnitude (\citealt{Coil09}, Figure 2).

\bigskip
Combining good redshifts from both surveys, out of 895 unique X-ray sources with secure optical counterparts, we have redshifts for 493 objects, including 25 stars and 468 galaxies. 
If we limit to the 426 sources with $I_{\rm AB}< 22$ (the limit of our main sample used in this paper), we have 354 secure redshifts out of 388 X-ray sources targeted, corresponding to a redshift success rate of 91\% and an overall completeness rate of 83\%.

The rest-frame $U-B$ colors of galaxies in both spectroscopic samples were derived using the $K$-correction code described by \cite{Willmer06}. Stellar masses were derived by \cite{Bundy06} from fitting spectral energy distributions to Palomar/WIRC $J$- and $K_s$-band photometry and CFH12K BRI photometry. For galaxies without \cite{Bundy06} stellar mass estimates, we substituted the stellar masses computed from absolute $M_B$ magnitude and rest-frame $B-V$ color using the prescription given by \cite{Bell03} and calibrated to the \cite{Bundy06} stellar mass scale with color- and redshift-dependent corrections \citep{Lin07, Weiner09}. 

\subsection{Emission-line measurements}\label{sec:emiline}

We measured the emission line fluxes in each spectrum after fitting and subtracting the stellar continuum. As in \cite{Yan06}, each spectrum was fitted by a linear combination of two templates after blocking the wavelengths corresponding to emission lines. The templates were constructed using the \cite{BC03} stellar population synthesis code. One template features a young stellar population observed 0.3 Gyr after a 0.1 Gyr starburst with a constant SFR. The other template is an old 7 Gyr simple stellar population. Both templates are modeled assuming solar metallicity and a Salpeter initial mass function. After subtracting the stellar continua, we simply summed the flux around the emission lines and divided by the median continuum level of the original spectrum (before subtraction) in two bracketing sidebands to get equivalent widths (EWs). The definitions for the central bands and sidebands are the same as in \cite{Yan06}. 

Emission line luminosity estimates require accurate flux calibration and correction for slit losses. Both are difficult to measure to better than 10\% accuracy. On the other hand, broadband photometry usually has much smaller errors. Using broadband photometry, we can apply $K$-corrections to get the rest-frame total flux in an artificial filter corresponding to our sidebands in EW measurements. Combining this with the EW measurements yields total line flux and luminosity, avoiding the need for spectrophotometric flux calibration \citep{Weiner07}. The accuracy of this method relies on the assumption that the emission line EW does not vary spatially within a galaxy. 
This assumption may not be accurate for AGNs, since the narrow-line regions subtend smaller scales than the stars in the host galaxies, leading to slightly larger \oiii\ EWs in the central regions of AGN hosts. However, for galaxies whose angular sizes are small compared to the seeing, the smearing of light by the atmosphere will make the emission line EW more uniform across the galaxy. 
X-ray sources are frequently massive galaxies. The median angular FWHM of our main sample of X-ray sources ($0.3<z<0.8$) in the $R$-band is $1\farcs27$, which is slightly larger than the slit width and the seeing. Therefore, this procedure will only mitigate the error caused by the slit loss but not eliminate it, and the emission line luminosity may be overestimated for AGNs. However, we expect the resulting bias and uncertainty in the emission-line luminosity to be far smaller than those in the X-ray luminosity or other uncertainties involved in the analysis, such as X-ray variability and unknown extinction corrections.

We used the DEEP2 CFH12K photometry catalog \citep{CoilNK04} and Blanton \& Roweis's (\citeyear{BlantonR07}) {\it k-correct} code v4\_1\_4 to derive the rest-frame absolute magnitudes for the sidebands around \hb\ and \oiii. We then converted the magnitudes to fluxes and multiplied them by the EWs to compute the total line fluxes and luminosities. 
The uncertainties for the emission-line EWs were scaled up according to the differences from repeated observations, following \cite{Balogh99} and \cite{Yan09}. The EW uncertainties are propagated into the luminosity uncertainties, along with uncertainties from photometry and $K$-corrections. Throughout the paper, we call an emission line {\it detected} if the EW is at least twice as large as its uncertainty. For non-detections, the luminosity upper limits are taken to be twice the uncertainties, i.e., 2$\sigma$.

\subsection{X-ray upper limits for optical sources}

Besides studying the X-ray selected sample, we have also investigated the X-ray properties of the optically selected sample. Therefore, we estimated the X-ray flux upper limits for sources not detected in X-ray. When extracting the X-ray counts at positions of the optical sources, we found that many of them have significant counts above the background. In many cases, the probability that the counts arise from background fluctuation is less than 0.1\%. The number of such cases is much higher than the expected number of false-positive cases, suggesting that many of them could be truly associated with the optical sources. It is very tempting to include them as X-ray detections to increase the sample size. However, because the X-ray PSF is usually larger than the PSF in the optical images, the X-ray flux could come from other untargetted galaxies nearby or even galaxies beyond the optical photometry detection threshold. Therefore, we only quote X-ray flux upper limits for these sources although the X-ray flux could be significant.

We estimate the X-ray flux upper limits using the Bayesian method \citep[][L09]{Helene84, Kraft91}. However, unlike L09, who used a power-law prior for the source flux distribution, we used a constant, non-negative prior \citep{Kraft91} with minimal assumptions made about the source flux distribution. This gives more conservative upper limits than the power-law prior. As we do not require an X-ray detection, we do not suffer from Eddington bias. Our upper limits correspond to the $95\%$ confidence limit ($2\sigma$).
In detail, the method applies the Bayes' theorem, 
\begin{equation}
P(s|N,b) \propto L(N|s,b) \pi(s),
\end{equation}
where $L(N|s,b)$ is the conditional probability of observing $N$ counts given the expected source counts $s$ and expected background counts $b$; it follows the Poisson distribution, 
\begin{equation}
L(N|s,b) = {(s+b)^N \over N!} e^{-(s+b)}.
\end{equation}
$\pi(s)$ is the prior probability distribution of the expected source counts; following \cite{Kraft91}, we adopt a step function for $\pi(s)$, which is a constant when $s>0$ and is zero otherwise.
Bayes' theorem yields the posterior distribution function, $P(s|N,b)$, for the expected source counts given the observed counts and the expected background. 

Here we assume that the error in the estimated background is small, thus $b$ is known. For each source, the total counts were estimated in an elliptical aperture that contains 70\% of the energy in the simulated PSF. The aperture was obtained using the PSF look-up table provided by L09. The mean background counts for each source were derived the same way as by L09 but were estimated for the aperture with EEF (enclosed energy fraction) of 70\% instead of the 90\% EEF aperture.
Given the observed counts and the expected background counts, we estimated the upper limit of the source counts using the formulae given in \cite{Kraft91}. The count limit was then converted to a flux limit and $K$-corrected by assuming a power-law photon index of $\Gamma=1.9$.

\section{A New Emission Line Ratio Diagnostic}

\subsection{The traditional method}

\begin{figure}
\begin{center}
\includegraphics[totalheight=0.35\textheight]{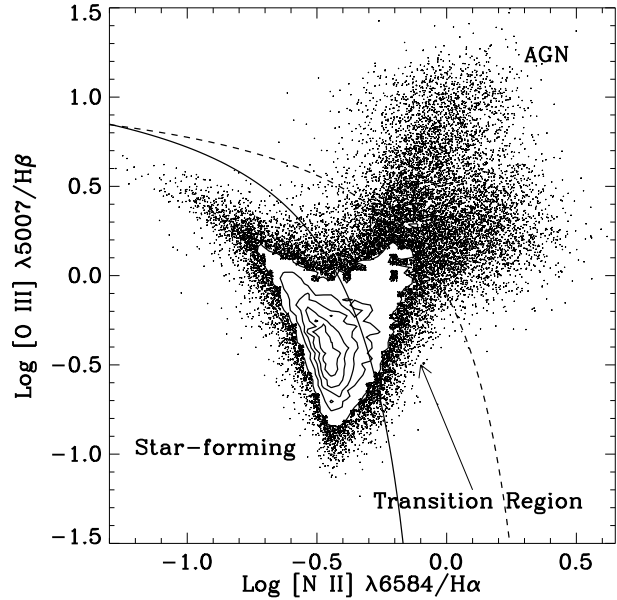}
\caption{Most commonly used line ratio diagnostic diagram (BPT) for a sample of the Sloan Digital Sky Survey (SDSS) galaxies with $0.05<z<0.1$, $r<19.77$, and all four emission lines detected at $>$$2\sigma$ level. The two curves indicate the AGN demarcations of \citet{KauffmannHT03} (solid line) and \citet{KewleyDS01} (dashed line). We use these demarcations to divide galaxies into three groups: AGN hosts (upper right), star-forming galaxies (lower left), and transition region (middle). Their distributions in our new diagnostic diagram are shown in Fig.~\ref{fig:sdsstest}.}
\label{fig:bpt_sloan}
\end{center}
\end{figure}

The classification of AGN and star-forming galaxies is usually achieved by the use of optical emission line ratio diagnostics (\citealt{BPT}, hereafter BPT; \citealt{VeilleuxO87}). One most commonly-used diagram involves two sets of line ratios: \niibw/\hal\ and \oiiibw/\hb\ (Figure~\ref{fig:bpt_sloan}). In such a diagram, a.k.a. the BPT diagram, the star-forming galaxies populate a sequence from the upper left to the lower center which is the consequence of a correlation between metallicity and ionization parameter. With increasing metallicity, the ionization parameter and hence \oiii/\hb\ decrease. Galaxies to the right and upper right of the star-forming sequence are AGN hosts, which include Seyferts ($\oiii/\hb\ > 3$) and low-ionization nuclear emission-line regions (LINERs, $\oiii/\hb\ < 3$). AGNs produce higher \oiii/\hb\ and \nii/\hal\ ratios than star-forming galaxies because of two reasons. First, AGNs have much higher ionization parameters so that they are capable of doubly ionizing more oxygen atoms. Secondly, their photons are more energetic than those from massive stars and can generate a more extended partially ionized region, where \nii\ is produced. 
The exact line ratios depend on the elemental abundances and the strength and shape of the ionizing spectra. For a more detailed discussion, see \cite{Stasinska06}. Nonetheless, the line ratio diagnostics provide a powerful way to distinguish AGN-dominated galaxies from star-formation-dominated galaxies.

Two empirical demarcations are commonly used for defining AGNs and illustrated in Figure~\ref{fig:bpt_sloan}. \cite{KewleyDS01} provided a demarcation based on extreme starburst models. It is a fairly conservative limit for defining AGNs. \cite{KauffmannHT03} proposed a more inclusive demarcation. Galaxies that lie between the two demarcations are often referred to as composite galaxies that have both AGN and star formation. Since SDSS fiber apertures include both the nucleus and the host galaxy, a lot of galaxies in this region are probably composites. However, this name is misleading in two ways. First, galaxies outside this intermediate region could also be composite galaxies. Second, some galaxies inside this region do not have to be composites, and there are evidences against the composite assumption. Using {\it HST}/STIS observations, \cite{Shields07} showed that in many such objects identified in the Palomar survey \citep{HoFS95}, the line ratios do not become more AGN-like on smaller apertures (10--20~pc), contrary to the expectation for composites that smaller apertures will include less star formation contribution.  Therefore, we choose to refer to the region in between the two demarcations as the ``Transition Region''. For the regions above and below, we refer to them as ``AGN'' and ``star-forming'', respectively. Note, our definition of ``Transition Region'' is different from that of the ``Transition Objects'' as defined by \cite{HoFS97III} based on a set of line ratios including \oi/\hal\ and using nuclear spectra. 

\subsection{Our new diagnostic method} \label{sec:newmethod}

As mentioned in \S\ref{sec:intro}, it is observationally very expensive to apply the traditional AGN diagnostics at $z>0.4$ due to the inaccessibility of \nii\ and \hal\ in the visible window. \cite{Weiner07} proposed a ``pseudo-BPT'' diagram using rest-frame $H$-band magnitude, $M_H$, to replace the \nii/\hal\ ratio. For star-forming galaxies, $M_H$, a proxy for stellar mass, correlates with the metallicity-indicating \nii/\hal\ ratio. Thus, this method can distinguish Seyferts in relatively massive hosts from low mass, low metallicity star-forming galaxies. However, because the correlation between stellar mass and \nii/\hal\ breaks down for AGN host galaxies, the separation between star-forming galaxies and AGNs is not very clean.
Inspired by \cite{Weiner07}, we propose a more effective classification method employing the optical $U-B$ color of galaxies in place of the \nii/\hal\ ratio. Below we first demonstrate this method using a galaxy sample from SDSS; then we explain why it works and why color works better than stellar mass.

\begin{figure*}
\begin{center}
\includegraphics[angle=90, totalheight=0.35\textheight, viewport=0 -10 380 760,clip]{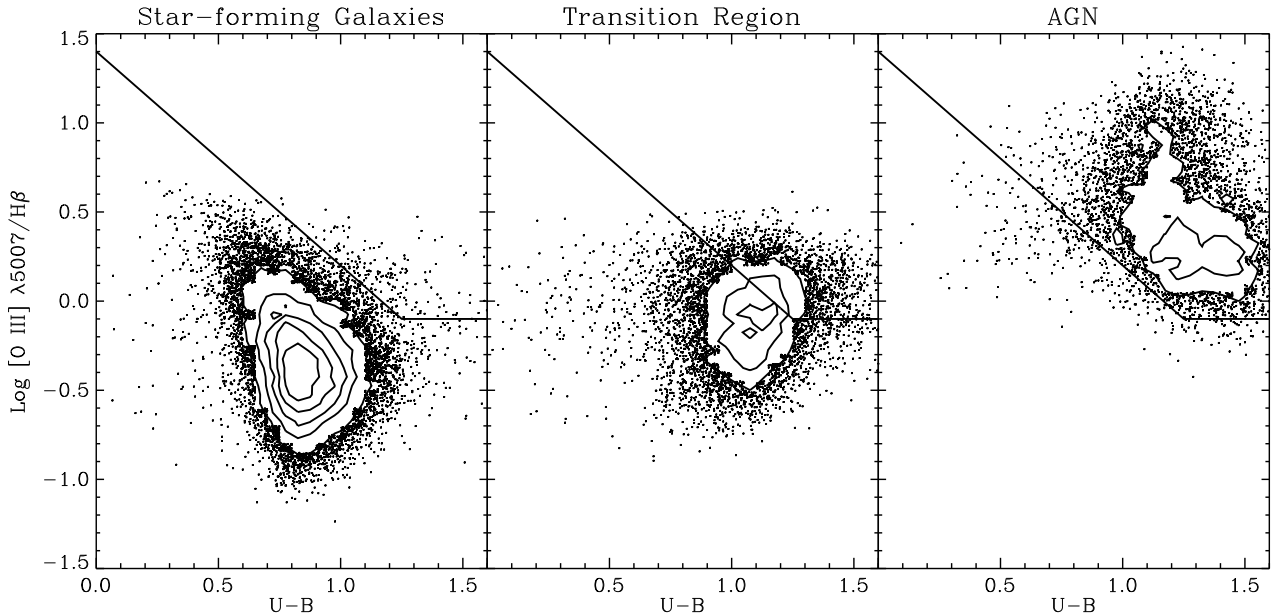}
\caption{Same sample of galaxies as in Fig.~\ref{fig:bpt_sloan}, but replacing \nii/\hal\ ratio on the horizontal axis with the rest-frame $U-B$ color. The three panels are for (a) star-forming galaxies, (b) transition region galaxies, and (c) AGN hosts, classified according to their positions in Fig.~\ref{fig:bpt_sloan}. The solid lines indicate our empirical demarcation for AGN selection, which removes nearly all pure star-forming galaxies (only 0.66\% of all star-forming galaxies in the sample cross this line).}
\label{fig:sdsstest}
\end{center}
\end{figure*}

The galaxy sample used in Fig.~\ref{fig:bpt_sloan} is selected from the Sloan Digital Sky Survey (SDSS, \citealt{York00}) --- Data Release Six \citep{SDSSDR6} by requiring $0.05<z<0.1$, $r<19.77$, and all four emission lines detected at more than $2\sigma$ significance. The emission lines are measured in the same way as by \cite{Yan06}. As discussed above, we separate this sample into three classes: star-forming galaxies, transition region galaxies, and AGNs. Figure~\ref{fig:sdsstest} replaces the horizontal axis with the rest-frame $U-B$ color.\footnote{The rest-frame $U-B$ color for SDSS galaxies was derived using \cite{BlantonR07}'s {\it k-correct} code v4\_1\_4.} The AGN hosts (panel c) are still in the upper right portion of the diagram, separated from the star-forming galaxies. The transition region galaxies overlap primarily with the star-forming galaxies in $U-B$ color. If we limit attention to pure AGNs, the $U-B$ color provides an effective alternative to the \nii/\hal\ ratio for selecting a sample. 





In choosing an empirical demarcation in the new diagnostic diagram, our preference is to limit contamination to the AGN sample. This new method certainly cannot select all galaxies containing AGNs down to the demarcation of \cite{KauffmannHT03} without being heavily contaminated by star-forming galaxies. We thus place the line just above the area populated by pure star-forming galaxies, which still allows us to retain most AGNs classified by the \cite{KewleyDS01} limit. The demarcation we use is given by

\begin{equation}
   \log (\oiii/\hb) > {\rm max}\{1.4-1.2(U-B) , -0.1\},
\label{eqn:demar}
\end{equation}

where max\{$a$,$b$\} denotes the greater value of $a$ and $b$, and $U-B$ is the rest-frame color in the AB magnitude system. This is illustrated by the lines in Fig.~\ref{fig:sdsstest}. 
The horizontal cut is a bit arbitrary: red galaxies with low \oiii/\hb\ ratio could be either transition region objects or very dusty star-forming galaxies. There are relatively few of them. We leave the fine tuning for future work.

With this demarcation (Eqn.~\ref{eqn:demar}), we find 95.7\% (7138 out of 7459) of the AGNs selected using the \citet{KewleyDS01} demarcation are still classified as AGNs using the new method. If we include all objects in the transition region as AGN hosts, the completeness of the new method drops to 54.3\% (9757 out of 17969). About 1.9\% (190 out of 9947) of the new ``AGNs'' were classified as star-forming galaxies under the old method; these we consider contamination. 


\subsection{The principle and the bias}




   

The new classification method is based on the fact that nearly all BPT-identified AGNs are found in red galaxies or those with intermediate colors between red and blue (hereafter, green galaxies), but hardly any are found in very blue galaxies. There are several reasons. First, blue galaxies are less massive and have smaller bulge-to-disk ratios than red galaxies. Smaller bulges host smaller black holes \citep{Magorrian98,McLure06}. The bluer a galaxy is, the less massive its black hole is, and at a fixed Eddington ratio, the less luminous the AGN will be. Thus, a lower fraction of the AGNs in blue galaxies will be found above the observational flux threshold than those in green or red galaxies. Second, star formation in blue galaxies could overwhelm weak or moderate AGNs so that the combined line ratios still put them in the star-forming sequence on the BPT diagram. Both of these effects tend to hide AGNs in blue galaxies. There may also be other physical effects that we do not yet understand. Nonetheless, this observational fact allows us to use host galaxy color to reproduce the BPT selection of AGNs. 


On the other hand, nearly all red/green galaxies that have high \oiii/\hb\ ratios are found to be AGNs. This is because high \oiii/\hb\ ratios require high ionization parameters which are produced in two types of sources: AGN and low-metallicity hot stars. Low-metallicity stars are only forming in very blue star-forming galaxies. Therefore, when limited to red or green galaxies, the high \oiii/\hb\ ratios have to be due to an AGN.

Therefore, the rest-frame $U-B$ color can be used to track AGN activity in a similar fashion as the BPT diagram, because it correlates positively with the bulge mass and metallicity, and correlates negatively with the star formation rate. Other colors or spectral index (e.g., $D_n(4000)$) could also be employed provided they satisfy these criteria.

What kind of AGNs will our method miss? First of all, this method is not intended to select broad-line AGNs (Type 1) in which the broadband color is not dominated by the host galaxy. Among narrow-line AGNs, one might worry that this method will miss those Seyferts with high $L_{\rm[O~III]}/M_{\rm BH}$ which are shown to live in blue star-forming galaxies \citep{Kauffmann07}. However, since our demarcation is tilted, we will not miss the majority of these Seyferts: their ``blue'' colors are redder than our limit. We tested this using our SDSS sample described in \S\ref{sec:newmethod}. We derived $L_{\rm[O~III]}/M_{\rm BH}$ following \cite{Kauffmann07} and selected the 5\% of AGNs (defined using the Kewley demarcation) with the highest values of $L_{\rm[O~III]}/M_{\rm BH}$, 77\% of these are still classified as AGNs in our new diagram. This fraction is lower than that for all AGNs since these galaxies do fall on the blue edge of the AGN sample. Nonetheless, our new method can recover the great majority of them. 


LINERs will be missed by this method if they live in blue galaxies, which will make them LINER-star-forming composites. On the BPT diagram, they will belong to the transition region. We will not be complete for this category. The fraction missed depends on where we put the demarcation and which method we regard as giving the ``correct'' classification for each galaxy, if it can be well defined. By lowering the demarcation to include more transition region galaxies, we would have more ``contaminations'' from the BPT star-forming sequence. However, some of these ``contaminations'' could also be AGN--SF composites. Regarding to SF--AGN composites, our method has a similar selection bias as the BPT diagrams, but with different detailed dependences.

Locally, we do not find low-metallicity AGNs \citep{Stasinska06}. If there exist at higher-$z$, they could be missed by both our methods and the traditional BPT diagram. 

There is also much evidence now suggesting that many luminous AGNs have previously had a star formation episode \citep{KauffmannHT03,Jahnke04,Sanchez04,Silverman08}, so they have bluer host galaxies than their inactive counterparts. Would we miss the AGNs in these galaxies? The answer is ``No''. Most post-starbursts in SDSS and DEEP2 have redder $U-B$ colors than the median star-forming galaxy \citep{Yan09}, because the $U$ band covers the blue side of the Balmer break. We therefore do not expect to miss AGNs in most post-starbursts, at least up to $z\sim0.8$.




The principal advantage of this method is that it requires fewer emission lines than the traditional one. Hence, it can be applied to higher redshift galaxies and suffers less from incompleteness due to missing line detections, especially in low signal-to-noise (S/N) spectra.

\subsection{Comparison to other methods} \label{sec:method_comp}

\begin{figure}
\begin{center}
\includegraphics[totalheight=0.35\textheight]{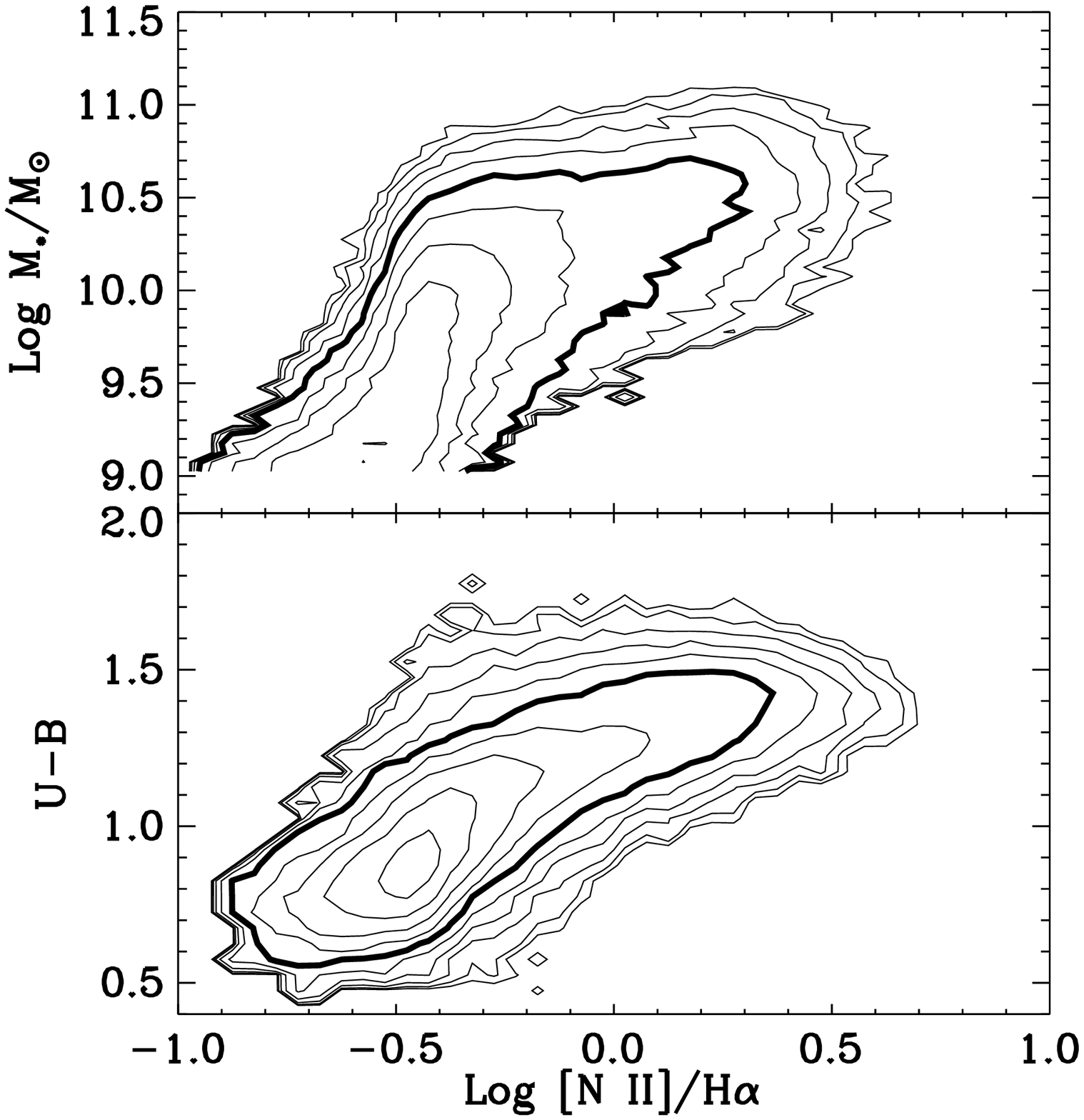}
\caption{Correlations between \nii/\hal\ and $M_*$ (upper panel) and between \nii/\hal\ and $U-B$ (lower panel). The sample is from the SDSS with selection described in \S\ref{sec:method_comp}. The contours denote equal density levels, $1/V_{\rm max}$ weighted so they reflect the distribution of a volume-limited sample. The contours are logarithmically spaced with each level inward representing a factor of two increment in number density. The highlighted thicker contour encloses 88\% (upper panel) or 90\% (lower panel) of all galaxies in a volume-limited sample.
}
\label{fig:mass_ub_comp}
\end{center}
\end{figure}

One might expect that stellar masses would perform equally well as a replacement for \nii/\hal.  However, this is not the case. Figure~\ref{fig:mass_ub_comp} shows how mass compares with rest-frame $U-B$ color in their correlations with \nii/\hal. The sample used is selected from SDSS DR6 with $0.02<z<0.1$, $M_* > 10^9 M_\odot$, and by requiring both \nii\ and \hal\ are detected at more than 2$\sigma$ significance. The stellar masses are derived as a by-product in $K$-correction \citep{BlantonR07}. We imposed a redshift-dependent stellar mass cut so that at all redshifts the sample is complete to a certain stellar mass limit for all colors. With this cut, for each galaxy, we computed the maximum volume over which a galaxy with that stellar mass would be included in the sample, $V_{\rm max}$. Figure~\ref{fig:mass_ub_comp} plots the $1/V_{\rm max}$-weighted distribution in \nii/\hal\ versus $M_*$ and \nii/\hal\ versus $U-B$ spaces. These reflect the distributions of all galaxies in a volume-limited sample down to $M_*>10^9M_\odot$ that have a reliable \nii/\hal\ measurement. 

In Figure~\ref{fig:mass_ub_comp}, the rest-frame $U-B$ color shows a much better correlation with \nii/\hal\ ratio than stellar mass does. The difference is especially dramatic at $\log (\nii/\hal) > -0.2$, where the trend for stellar mass becomes horizontal.
These high \nii/\hal\ galaxies are mostly AGN-host galaxies that are old and have very low or zero SFRs (as indicated by their red colors).
Both \nii/\hal\ and color increase with metallicity in passively evolving galaxies (see \citealt{GrovesDS04II} for the dependence of \nii/\hal\ on metallicity for AGNs), and hence they remain correlated as metallicity changes. However, although stellar mass correlates with metallicity for star-forming galaxies, stellar mass stops growing once the star formation stops. Thus, it no longer correlates with the \nii\ abundance or \nii/\hal\ ratio. Therefore, color provides a better alternative to \nii/\hal\ than stellar mass does.

Lamareille et al. (\citeyear{Lamareille04, Lamareille10}) investigated the use of \oiiw/\hb\ EW ratio as an alternative to \nii/\hal. This enables the application to redshifts as high as our method, though it also requires that \oii\ lines are covered in the spectra. This method also has troubles differentiating the transition region from the star-forming galaxies. Under their method, 84.5\% of all Seyferts and LINERs (above the \citealt{KewleyDS01} demarcation) are classified as AGNs, smaller than our completeness of 95.7\%. The contamination from the star-forming sequence (below the \citealt{KauffmannHT03} demarcation) to the AGN category is 2.6\%, larger than our number of 1.9\%.

\subsection{Intermediate-$z$ test}

The SDSS galaxy sample we used is at $z<0.1$. We have shown above that the new method works at this redshift. Does it still work at higher redshifts? 

\begin{figure*}
\begin{center}
\includegraphics[totalheight=0.35\textheight]{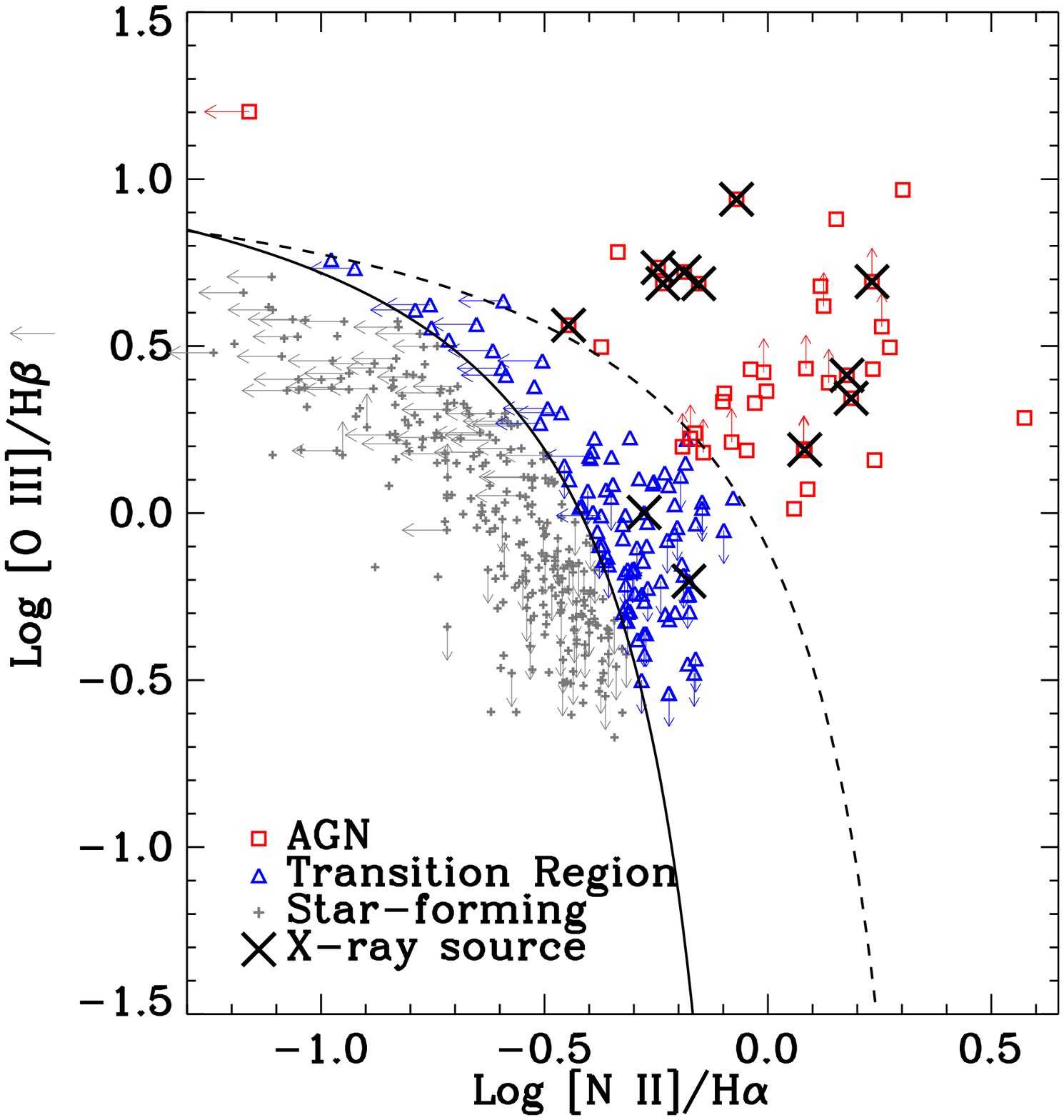}
\includegraphics[totalheight=0.35\textheight]{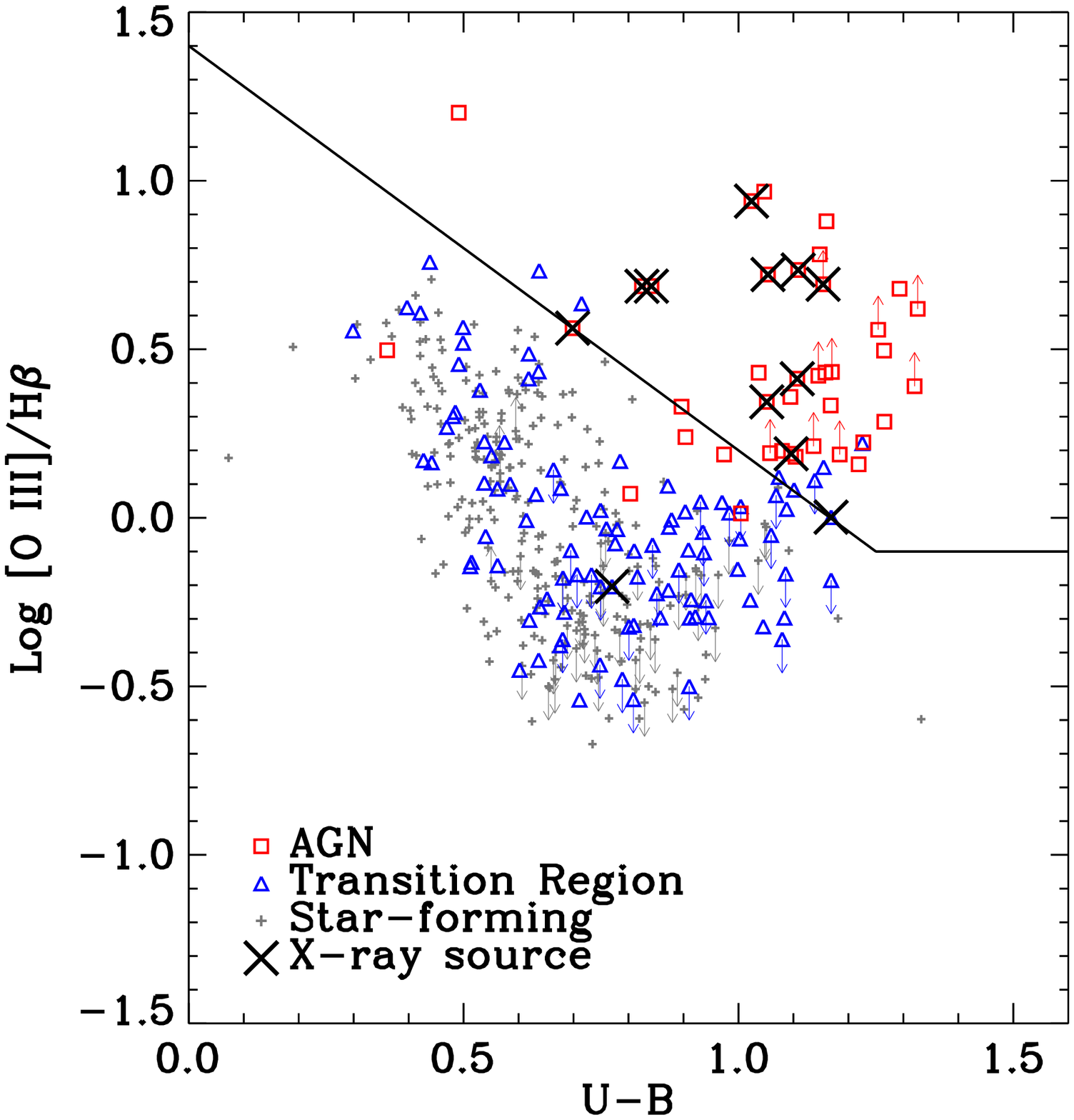}
\caption{Left: standard line ratio diagnostic diagram for a sample of sources in the EGS with $0.2<z<0.4$. Arrows indicate the 2$\sigma$ upper and lower limits for galaxies in which one of the four lines is not significantly detected. The solid and dashed curves show the demarcation used by \citet{KauffmannHT03} and \citet{KewleyDS01} to separate AGN hosts from star-forming galaxies. We use the two curves to classify all galaxies into three categories: red squares indicate AGN host galaxies; small gray crosses are star-forming galaxies; and blue triangles are galaxies in between, which is usually considered as composite objects. Large dark crosses indicate sources detected in the X-ray. Right: the same galaxies now plotted on the $U-B$ vs. \oiii/\hb\ diagram. The AGN hosts are still at the upper right portion, separated from star-forming galaxies. The solid lines mark our empirical cuts, which were designed on the basis of lower-redshift, SDSS data.}
\label{fig:egstest}
\end{center}
\end{figure*}

We can test our new AGN/star-forming classification method between redshifts 0.2 and 0.4, where the traditional line ratio diagnostics are still available within the optical window. We used the spectroscopic data obtained in the EGS by the DEEP2 survey and with Hectospec to test the method. The results are shown in Fig.~\ref{fig:egstest}. 
The traditional method identified 40 emission line AGNs that are above the \cite{KewleyDS01} demarcation.\footnote{For four objects near the demarcation with arrows pointing across it, the limits on line ratios strongly suggest that they belong to the category across the demarcation. Thus, we assigned them those classifications.} The new method identified 36 of them and missed 4, corresponding to a 90\% completeness. It also picked up 8 objects from the transition region with no contamination from star-forming galaxies. It is also encouraging to note that as about many X-ray sources (11 out of 12) are identified as AGNs in the new diagram as in the traditional diagram (10 out of 12). 

Because galaxies are bluer at higher redshift \citep{Blanton06}, in principle, our demarcation should shift blueward slightly. By comparing the color--magnitude diagram of the DEEP2 sample at $z\sim0.9$ with that of an SDSS galaxy sample at $z\sim0.1$, we found the division between red and blue galaxies shifts by 0.14 mag in $U-B$ between these redshifts \citep{CooperNW08, Yan09}, which is consistent with the passive evolution prediction \citep{vanDokkum01}. The corresponding shift between $z=0.3$ and $z=0.1$ is about 0.04. For the sample we will discuss later, which has $0.3<z<0.8$ and a median redshift of $z\sim0.55$, the shift is about 0.08 from $z\sim0.1$. These shifts are quite small and insignificant for our results. Considering our sample covers a wide redshift range, for simplicity, we do not apply these shifts.  


\section{X-ray Selection vs. Optical Selection}

A commonly used method to identify X-ray AGNs is to use a pure luminosity cut of $L_{2-10 {\rm keV}}> 10^{42}~{\rm erg~s^{-1}}$. 
This is a very conservative threshold which is based on the fact that no local star-forming galaxies have X-ray luminosity above it \footnote{Using the calibration by \cite{Ranalli03}, one would need an SFR of $200 M_\odot~{\rm yr}^{-1}$ to produce enough X-ray luminosity from non-AGN sources to cross this threshold. }. However, lower luminosity sources could also be bona fide AGNs, which are equally, if not more, interesting. The advantage of a selection based on X-ray luminosity is its rough correspondence to a bolometric luminosity selection \citep{Elvis94}. However, this can be compromised by intrinsic absorption of the X-ray luminosity. Below, we compare the X-ray selection to optical emission line selection in the DEEP2 and Hectospec sample. 

\begin{figure}
\begin{center}
\includegraphics[totalheight=0.35\textheight]{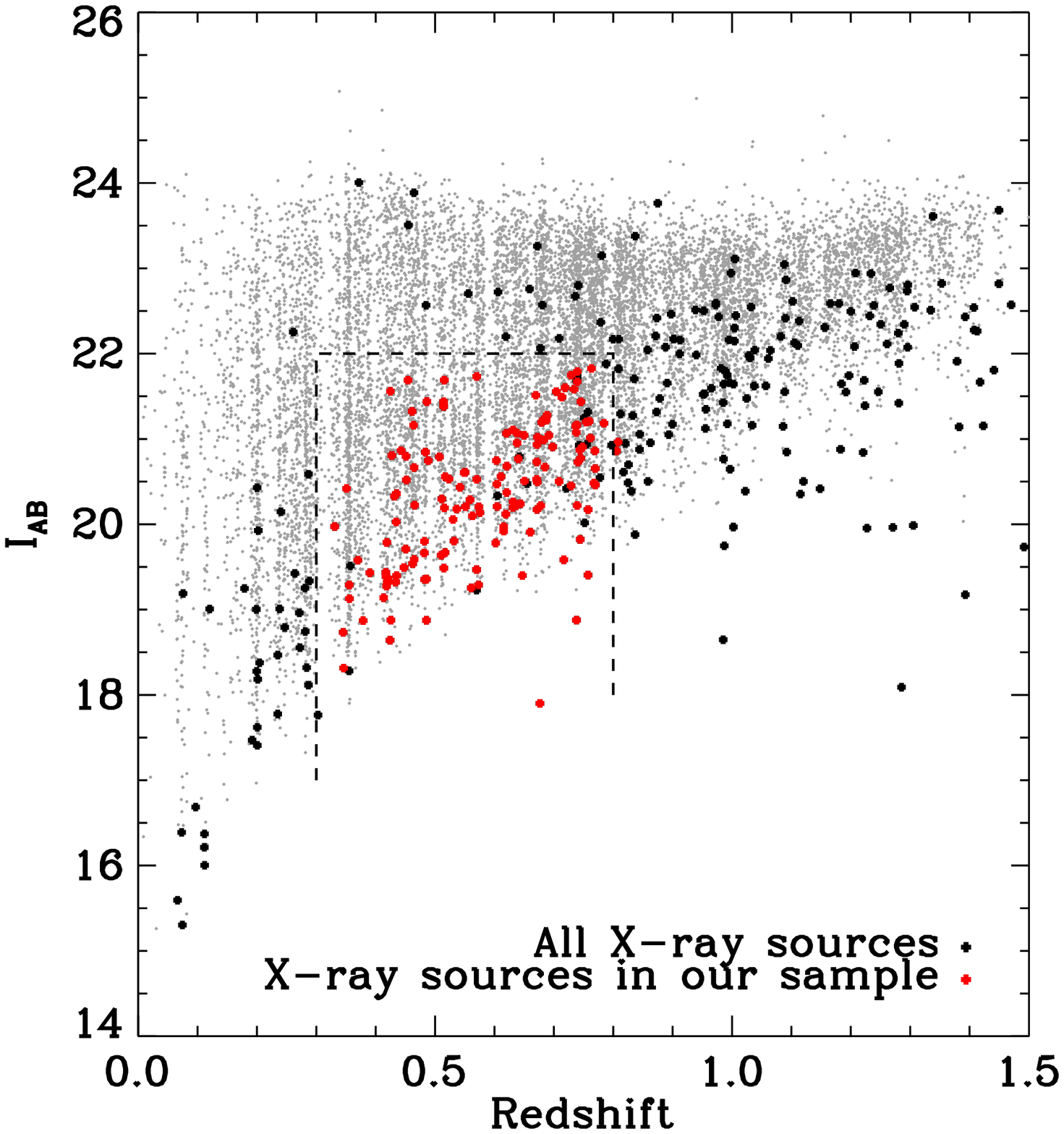}
\caption{Magnitude and redshift distribution of X-ray sources (large points) that have optical counterparts and successful redshifts from the DEEP2 survey and/or the Hectospec follow-up program, compared with the overall distribution of galaxies (small gray points) with redshift successfully obtained in these two surveys. The red points are X-ray sources in our sample. The dashed lines indicate the magnitude limit and the approximate redshift limits. The latter is approximate due to the slightly varying wavelength coverages of DEEP2 spectra. Objects are rejected from the sample if their spectra do not cover both \oiii\ and \hb, which could also be due to CCD gaps, bad columns, etc. We also exclude all objects at $z<0.3$.}
\label{fig:x_zdist}
\end{center}
\end{figure}

\begin{figure}
\begin{center}
\includegraphics[totalheight=0.35\textheight]{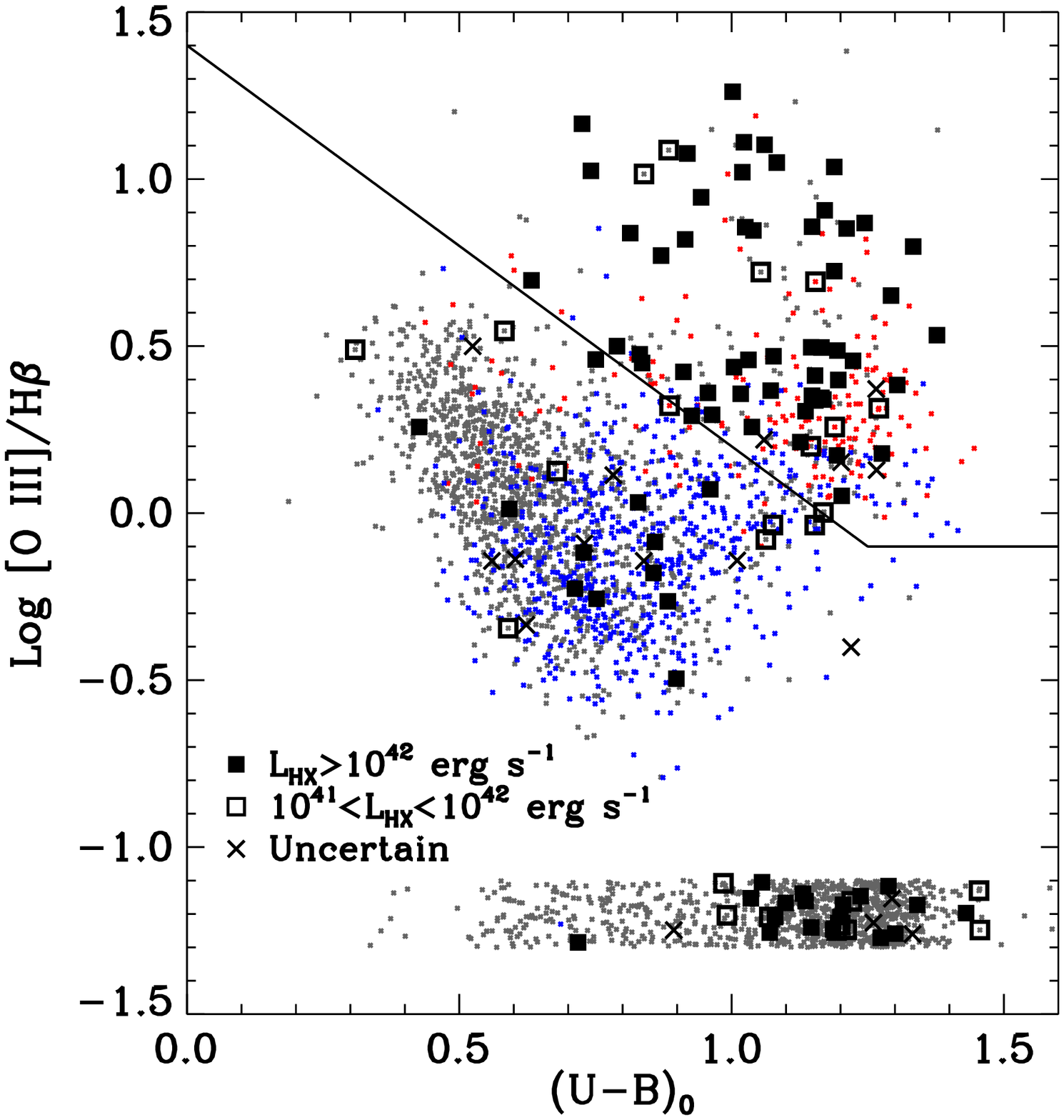}
\caption{New emission-line diagnostic diagram for DEEP2 galaxies with $0.3<z<0.8$. The solid line shows the proposed demarcation between star-forming galaxies (below and to the left) and AGNs (above and to the right). Galaxies with either \oiii\ or \hb\ undetected are placed at their 2$\sigma$ upper (blue points) or lower (red points) limits, respectively (arrows are omitted for clarity). In nearly all cases, these line ratio limits do not affect the classification of objects. Quiescent galaxies without detectable emission lines  are also shown at the bottom. X-ray sources, which are found in all three categories, are marked as large symbols. Solid squares are bright sources with $L_{2-10{\rm keV}}>10^{42} {\rm erg~s^{-1}}$; open squares are faint sources with $L_{2-10{\rm keV}}<10^{42} {\rm erg~s^{-1}}$; and crosses are X-ray sources with uncertain luminosity classes---they are usually undetected in the hard band. Only objects in the first category may be definitely classified as AGNs, though many in the second class will be AGNs.} 
\label{fig:ub_o3hb_xdet_lum}
\end{center}
\end{figure}


We limited our sample to objects that have both \oiii\ and \hb\ covered in the spectra, which corresponds roughly to a redshift range of $0.3<z<0.8$. The redshift range is approximate due to the slightly varying wavelength coverage of the DEEP2 spectra. We made a magnitude cut at $I_{\rm AB}<22$ so that our redshift success rate is above 90\% for both red and blue galaxies, and the spectra have a sufficient S/N for stellar continuum subtraction. 
As shown in Figure~\ref{fig:x_zdist}, few X-ray sources at $0.3<z<0.8$ are excluded by this cut. To summarize, all sources in our sample have to satisfy all of the following criteria: 

\begin{enumerate} \itemsep0pt
\item be within the X-ray footprint;
\item $I_{\rm AB}<22$;
\item have reliable redshifts from either the DEEP2 survey or the Hectospec follow-up;
\item have \oiii\ and \hb\ well covered in the spectra, not badly affected by CCD gaps or very bright sky lines;
\item $z>0.3$.
\end{enumerate}
In total, there are 3150 galaxies and 146 X-ray sources in this sample. 


In our analysis, we primarily focus on Type 2 AGNs since their comoving number density is much higher than that of Type 1 AGNs, and the latter are usually identified easily in both optical spectra and X-ray data. However, we will use Type 1 AGNs as a reference sample. We visually identified Type 1 AGN candidates among the spectra from DEEP2 and the MMT/Hectospec follow-up survey. We measured the FWHM of the broad lines (\hal, \hb, or \mgii) and classified those with FWHM greater than 1000~km~s$^{-1}$ as Type 1 AGNs. There are 21 Type 1 AGNs in our sample. All but one are detected in the X-ray. The one undetected object is not far from the detection threshold in the hard band, with a false-positive probability of $1.8\times10^{-3}$. 

In Figure~\ref{fig:ub_o3hb_xdet_lum}, we present the new emission-line diagnostic diagram for all non-Type-1 galaxies in our sample. The distribution is similar to that at $0.2<z<0.4$ (right panel of Fig.~\ref{fig:egstest}). For galaxies with either \oiii\ or \hb\ undetected, we place them at their lower or upper limits for \oiii/\hb. In 89\% of such cases (726 out of 815), the upper limits on the undetected lines are tight enough that they do not introduce any ambiguity in the classifications of the objects. The remaining 11\% are classified as ``ambiguous'', which represents 3\% of our sample and can be safely neglected. 
 Figure~\ref{fig:ub_o3hb_xdet_lum} also plots those galaxies with neither \oiii\ nor \hb\ detected at the bottom. Effectively, we classify all galaxies into three main categories: star-forming, AGNs, and quiescent. 
The X-ray detected sources (excluding Type 1 AGNs) are found in all three categories. We use solid symbols to denote sources with $L_X(\hard)>10^{42} {\rm erg~s^{-1}}$ and open symbols to denote fainter sources. The fainter sources either have a detected $L_X(\hard) > 10^{41} {\rm erg~s^{-1}}$, or, in the cases of hard band undetection, have a minimum $L_X(\hard)$ extrapolated (assuming $\Gamma \leq 1.9$) from the \soft\ band greater than $10^{41} {\rm erg~s^{-1}}$.

The differences between the two selection methods are apparent in this figure. First, consider the optically selected Type 2 AGNs: these are points above the demarcation. Many of them (78\% of all optically selected Type 2 AGNs) are not detected in X-rays. Some are detected but are fainter than the commonly used $10^{42} {\rm erg~s^{-1}}$ threshold. The majority (51\%) of non-Type-1 X-ray sources with $L_{X} (\hard) > 10^{42} {\rm erg~s^{-1}}$ are also optically classified as AGNs. However, 22\% of them are found in the star-forming part of the line ratio diagram and 25\% are found to have no detectable line emission. As mentioned in \S\ref{sec:intro}, these cases are often referred to as XBONGs or optically-dull X-ray galaxies. Often, the term XBONG is used to refer both to galaxies with no detectable line emission and to those with line ratios of typical star-forming galaxies. We suggest treating these two cases separately as the galaxies are two distinct types; in the remainder of this paper, we only use XBONG to refer to the class with no detectable line emission, and thus our XBONGs are nearly all red-sequence galaxies.

To understand the reason for the discrepancy between optical selection and X-ray selection, we investigate a few classes of objects, grouped according to the differences in X-ray and optical classifications. To simplify things, we do not separate the bright and faint X-ray sources into two separate categories, since the luminosity threshold is somewhat arbitrary. We have three optical classifications---AGN, SF, and quiescent---and two X-ray classifications---detections and non-detections. 

First, we will consider the objects that are classified as AGNs by the optical selection and are detected in X-ray. For brevity, we will call them ``unambiguous AGNs''. 
Second, we will discuss X-ray detected sources that are found in the star-forming branch of our diagnostic diagram. Following \cite{Moran96} and \cite{Levenson01}, we call them ``X-ray-loud composite galaxies'', reflecting their AGN-star-forming composite nature, as we will show below. Third, we will investigate X-ray sources with no detectable line emission (XBONGs). Lastly, we will discuss optically selected AGNs with no X-ray detection, which we refer to as ``optical-only AGNs''. We will use these names for each class throughout the remainder of this paper. Their definitions are summarized in Table \ref{tab:classes}, along with the number of objects in each class. Note, because of the preferential follow-up of the X-ray sources, the numbers in this table should not be used to estimate the fraction of emission-line AGNs that are detected in the X-ray. For that analysis, we limit the sample to only sources targeted in the DEEP2 survey, where no preference was given to X-ray sources.

As will be shown below, the union of X-ray selected AGNs and optically selected AGNs provides a much more complete AGN sample. Table.~\ref{tab:xobj_prop} lists the IDs, coordinates, redshifts, optical colors, emission line properties,  and the classifications of all the X-ray sources and the optical-only AGNs in our sample, along with Type 1 AGNs. 

\begin{table*}
\begin{center}
\caption{Classification Definitions for Non-Type-1 Objects}
\begin{tabular}{ccc}
\hline \hline
Emission Line (3129) &	X-ray Detected (126) & X-ray Undetected  (3003) \\\hline
AGN (291)      &  Unambiguous AGNs (64) &   Optical-only AGNs  (227) \\
SF (1799)      &  X-ray-loud composite galaxies (28) &      \\
Quiescent (950) &  XBONGs (32) &   \\
Ambiguous (89) &  Ambiguous (2) &  \\\hline
\end{tabular}
\tablecomments{The numbers in parentheses indicate the sample size in each class. 
}
\label{tab:classes}
\end{center}
\end{table*}


\subsection{Unambiguous AGNs}

\begin{figure*}
\begin{center}
\includegraphics[totalheight=0.35\textheight]{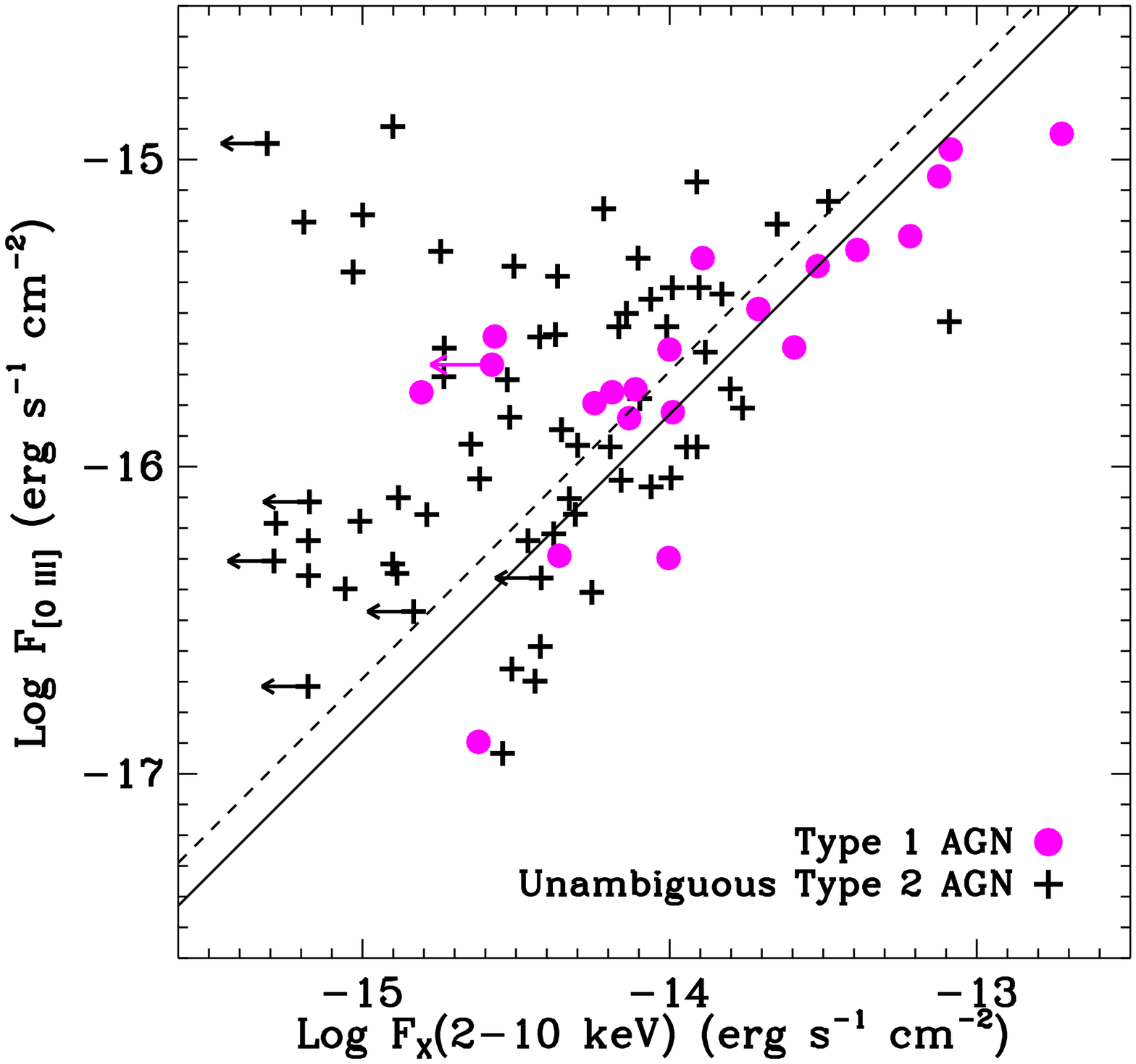}
\includegraphics[totalheight=0.35\textheight]{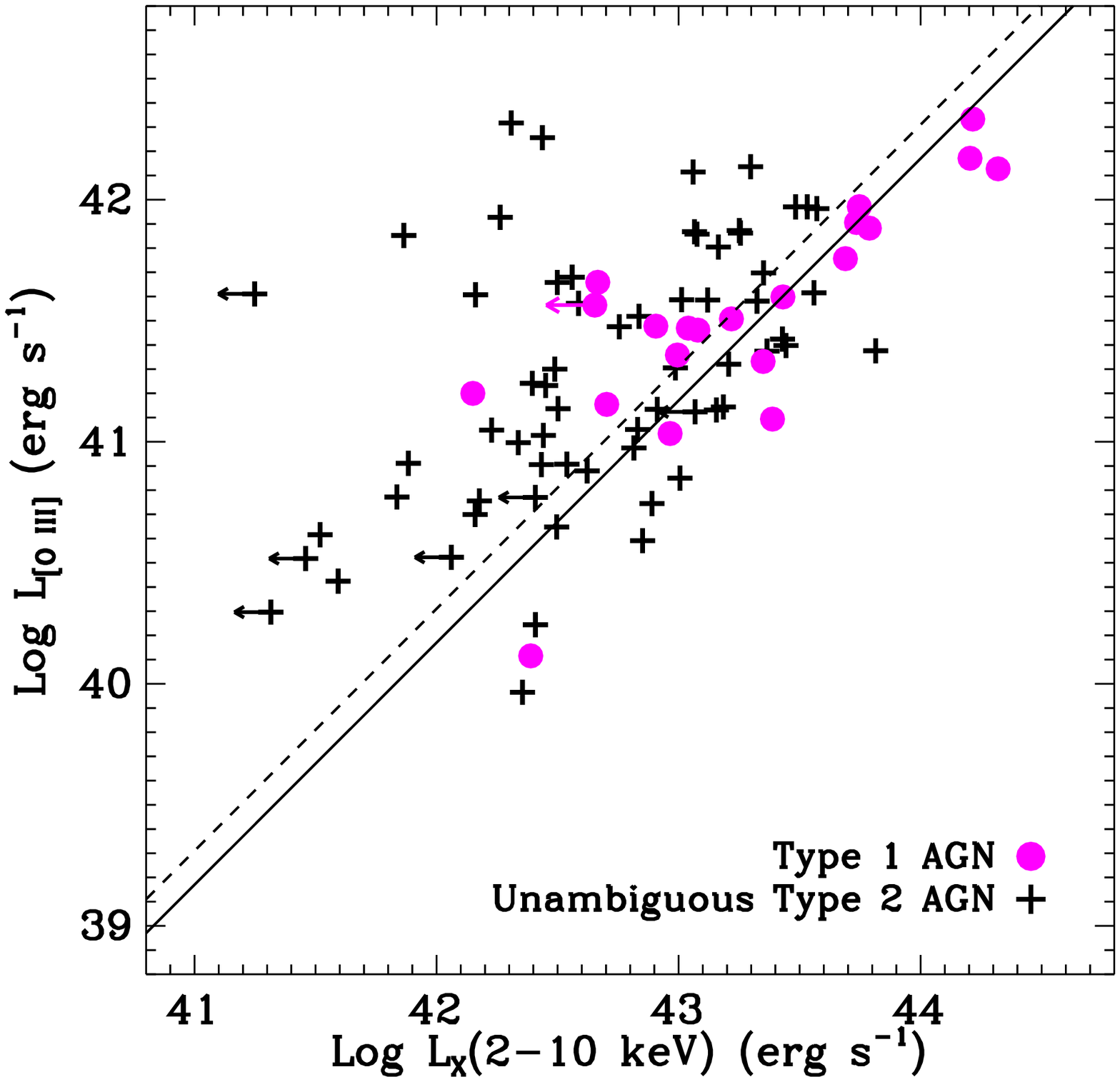}
\caption{Left: \oiii\ line flux versus 2--10~keV X-ray flux for
  unambiguous AGNs, which are X-ray sources that are classified as AGNs
  by emission line diagnostics. Type 1 AGNs are also shown here as
  solid magenta circles. The dashed line indicates the median ratio
  found by \cite{Heckman05} for Type 1 AGNs. The solid line indicates
  the median ratio in our Type 1 sample. Right: similar to the left 
  panel but comparing luminosity rather than observed flux.} 
\label{fig:o3_x_class1}
\end{center}
\end{figure*}

\begin{table*}
\begin{center}
\caption{Median $\log L_X(\hard)/L_{\rm[O~III]}$ Ratios for Unambiguous AGNs}
\begin{tabular}{cccc}
\hline \hline
Class     & HX-selected & \multicolumn{2}{c}{\oiii-selected} \\
          & Our Sample  & \cite{Heckman05} & \cite{LaMassa09} \\ \hline
Type 1 & 1.83 (0.29) & 1.69 (0.41) &\\
Type 2 & 1.42 (0.50) & 0.68 (1.16) & 0.46 (1.01) \\
Combined & 1.57 (0.53) & 1.22 (1.01) &\\
\hline
\end{tabular}
\tablecomments{Numbers in parentheses are the scaled median absolute
  deviations $S_{\rm MAD}$ of each  sample.  Median absolute deviation (MAD) is a robust
  estimator of distribution width for small samples. We follow \cite{Beers90} to define
  $S_{\rm MAD} = {\rm MAD}/0.6745$, which is 1 for a normal distribution 
  with ${\rm standard~deviation} =1$.  Statistics from
  \cite{Heckman05} and \cite{LaMassa09} have been computed from their data tables.}
\label{tab:ratios}
\end{center}
\end{table*}

Most (83\%) optically classified AGNs that are also detected in X-rays
have $L_X(\hard) > 10^{42}$~erg~s$^{-1}$,
which confirms their identity as AGNs.  The
correlations between the emission line and X-ray luminosities of this
population establish prototype relations for AGNs.  Both \oiii\ and hard X-ray are
good indicators for AGN bolometric luminosity (\citealt{Heckman04b} for \oiii; \citealt{Elvis94} for X-ray). Because \oiii\
originates from the narrow-line region, which is outside the obscuring
dusty torus, it is usually regarded as an isotropic luminosity
indicator. While X-rays can penetrate dust easily, they can be
absorbed by a high column density of neutral gas in the torus. Among
Type 1 AGNs, for which we have an unobstructed view of the accretion
disk, X-ray luminosity is found to correlate with \oiii\ luminosity
\citep{Mulchaey94,Heckman05}. Therefore, comparing X-ray with \oiii\
can reveal the level of X-ray absorption
\citep{Maiolino98,Bassani99}. 

Figure~\ref{fig:o3_x_class1} compares \oiii\ emission with X-ray in both flux and
luminosity.  Type 1 AGNs have a slightly larger median
$\log (L_{X}(\hard)/L_{\oiii})$ ratio
(1.83 dex) and a narrower distribution than our Type 2 AGNs (median=1.42~dex, see Table~\ref{tab:ratios}). However, the $L_{X}(\hard)/L_{\oiii}$ difference
between Type 1 and Type 2 AGNs depends on how the sample is
selected. 
\citet{Heckman05} showed that in a hard-X-ray-selected sample of local
AGNs, Type 1 and Type 2 AGNs exhibit $L_{X}(\hard)/L_{\oiii}$ ratios
indistinguishable from each other.  
However, in a sample selected by \oiii\ luminosity, many Type 2's can be
present with low X-ray luminosity, presumably due to absorption. The Type 1 
AGNs then have a median ratio much larger than that of Type 2 and
have significantly less variation in the ratio than Type 2 AGNs. The unambiguous
AGNs are effectively a hard-X-ray-selected sample; there is a small
difference between Type 1 and Type 2 but not nearly as large as the
difference in an \oiii-selected sample.  Table~\ref{tab:ratios} compares the median and
distribution widths for the various samples. 

\subsection{X-ray-loud Composite Galaxies}

Thirty X-ray sources have
\oiii/\hb\ ratios and $U-B$ colors that place them in the 
star-forming area of the emission-line diagnostic diagram.
X-rays can be produced in star-forming galaxies by high-mass X-ray
binaries, low-mass X-ray binaries, supernova remnants, and hot
interstellar medium heated by supernova \citep{Fabbiano89} in addition
to possible AGNs. 
Many authors have shown that in starburst galaxies without an AGN,
the total X-ray luminosity correlates with the star formation rate
\citep{Nandra02,Bauer02,Ranalli03,Grimm03,Colbert04,Persic04,Hornschemeier05,
  Georgakakis06, Persic07, Rovilos09}.  Thus the expected X-ray flux
from the sources related to SF can be predicted if the SFR is known. 
We used the \cite{Ranalli03} calibration to estimate the expected
X-ray luminosity from sources related to star formation.

\begin{figure*}
\begin{center}
\includegraphics[angle=90, totalheight=0.35\textheight, viewport=30 0 500 680]{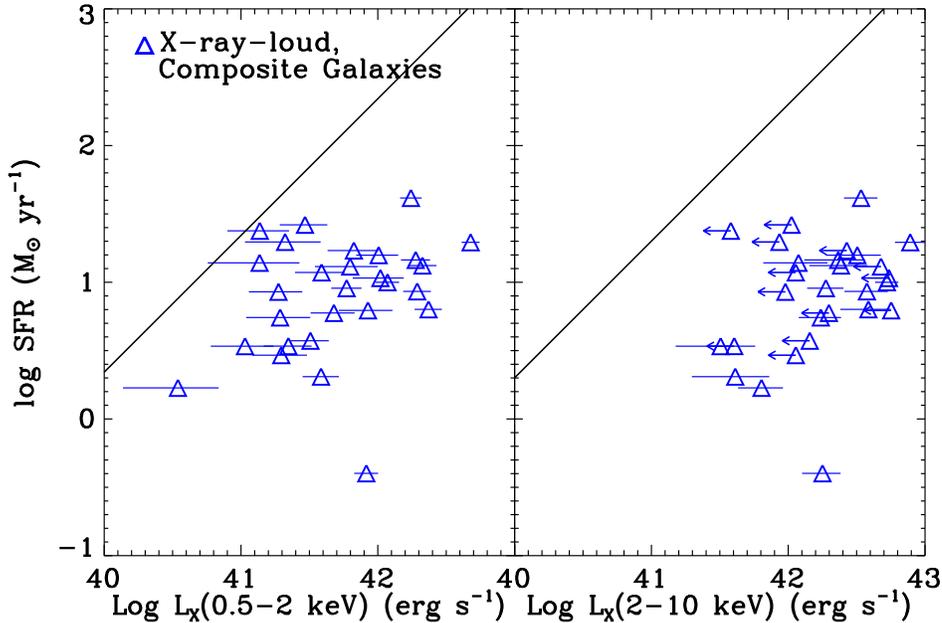} 
\caption{SFR derived from \hb\ versus the X-ray luminosity 
  for our X-ray-loud composite galaxies---i.e., X-ray sources that
  appear as star-forming galaxies in emission-line diagnostics. The
  left panel is for the soft band (0.5--2keV), and the right panel is
  for the hard band (2--10~keV). The solid lines indicate the SFR--$L_X$
  relation calibrated by \citet{Ranalli03}. Median uncertainty in $\log
  ({\rm SFR})$ is 0.20~dex and is dominated by the scatter in the
  level of extinction among galaxies.} 
\label{fig:hbsfr_x}
\end{center}
\end{figure*}

Figure~\ref{fig:hbsfr_x} compares observed X-ray luminosities with
star formation rates computed from \hb.  \cite{Moustakas06} provide
an empirical calibration to derive SFR from the observed \hb\ strength. 
The calibration coefficients depend on the rest-frame $B$-band absolute
magnitudes ($M_B$) of the galaxies.
We linearly interpolated between the points given in
Table~1 of \citet{Moustakas06}. This is equivalent to applying an
average extinction correction in bins of $M_B$. As shown by
\citet{Moustakas06}, when lacking a reliable extinction measurement
from \hal/\hb\ ratio, this empirical \hb\ calibration can achieve a
SFR estimate good to $\pm40\%$~(1$\sigma$). 

Most of
the X-ray-loud composite galaxies have X-ray luminosities much higher
than star formation can account for in both the soft  and
hard bands.  The excess is more than two orders of magnitude in the extreme
cases. A large fraction of the X-ray luminosity in these objects must
come from a central AGN\null.  These galaxies therefore appear to be
undergoing both star formation and nuclear activity. Since AGN will also contribute to the total \hb\ luminosity, the SFR could be overestimated, which leads to an underestimate of the AGN component.

\begin{figure}
\begin{center}
\includegraphics[totalheight=0.58\textheight,viewport=45 0 560 820,clip]{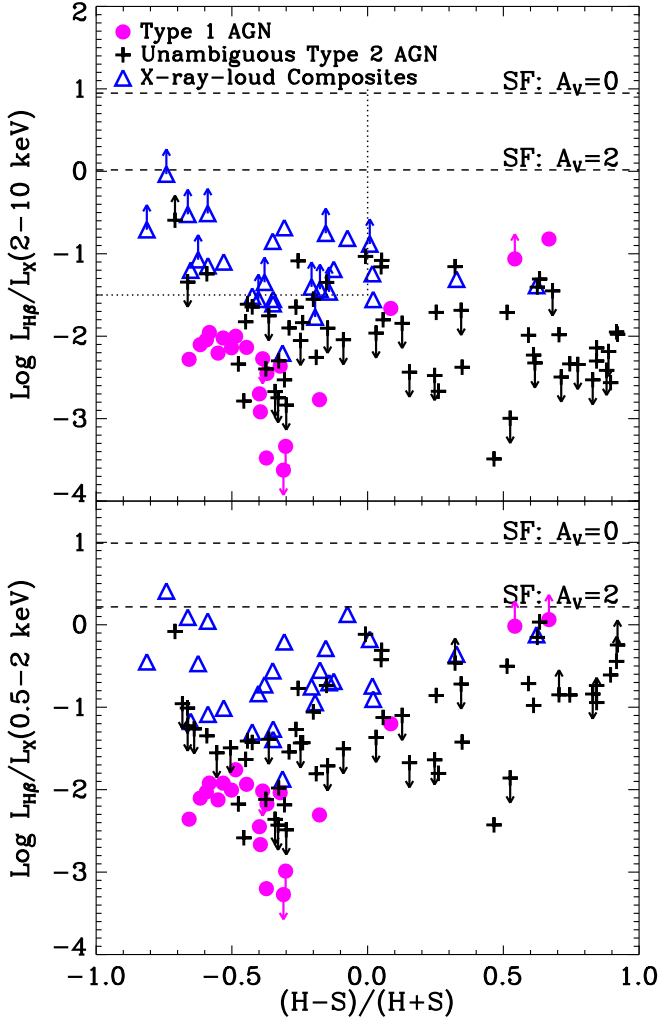}
\caption{Upper: $L_{\hb}/L_X$ versus hardness ratio for Type 1 AGNs (magenta
  circles), unambiguous Type 2 AGNs (black crosses), and X-ray-loud
  composite galaxies (blue triangles).  The two horizontal dashed
  lines mark the expected $L(\hb)/L_X$ ratios from star-forming
  galaxies with zero or 2 mag of extinction. The dotted
  lines indicate a rough demarcation for the boundary between the
  region occupied by X-ray-loud composite galaxies and typical
  AGNs. Here $L_{\hb}$ has not been corrected for extinction. The
  figure shows all Type 1 AGNs and all unambiguous Type 2 AGNs that
  are detected in either 2--10~keV band ($p<0.01$) or \hb. For objects 
  detected
  in one of these measures, 2--10~keV or \hb\ but not both, the
  corresponding upper or lower limits in the ratio are indicated by
  the downward or upward arrows, respectively.
Lower: same as upper but for the soft band
  (\soft). } 
\label{fig:hboverx_hr}
\label{fig:hboversoftx_hr}
\end{center}
\end{figure}

Further evidence for the coexistence of SF and a central AGN comes
from the distribution of the $L_{\hb}/L_X{(\hard)}$ ratio
as a function of the observed X-ray hardness ratio (HR), plotted in
Figure~\ref{fig:hboverx_hr}. 
X-ray-loud composite galaxies have $L_{\hb}/L_X{(\hard)}$ ratios
higher than typical AGNs but lower than pure star-forming
galaxies. Based on the \cite{Ranalli03} relation between $L_X{(\hard)}$
and SFR, and \cite{Kennicutt98}'s relation between $L_{\hal}$ and
SFR, assuming case B Balmer decrement $\hal/\hb=2.85$, typical
star-forming galaxies should have $L_{\hb}/L_X(\hard)$ greater than
1 (assuming $A_V\le2$). Unambiguous and Type 1 AGNs have an
$L_{\hb}/L_X(\hard)$ lower by two orders of magnitude, around
$10^{-2}$. However, most X-ray-loud composite galaxies have
intermediate ratios in $L_{\hb}/L_X(\hard)$. The simplest
explanation is that they are composite objects having both star
formation and active nuclei. Most of their \hb\ emission originates
from star-forming \hii\ regions, while most X-ray emission originates
from matter around the SMBH.  
For example, assuming the intrinsic $L_{\hb}/L_X(\hard)$ ratio for
AGNs is $10^{-2}$ and for pure star-forming galaxies is 10,  a
galaxy with $\log L_{\hb}/L_X(\hard)=-1$, in the absence of
extinction, 90\% of  the \hb\ emission
comes from star-forming \hii\ regions, and 10\% comes from the
narrow-line region around an AGN\null. In contrast, 1\% of the hard X-ray emission
comes from X-ray binaries and supernova remnants, and 99\% comes from
the AGN\null. If extinction on \hb\ is present at the same level 
for  both the star-forming and nuclear regions, the
resulting proportions  do not change.  
Figure~\ref{fig:hboverx_hr} shows a relatively clean separation
between unambiguous AGNs and X-ray-loud composite galaxies in the
$L(\hb)/L_X$ versus hardness ratio diagram. This supports the hypothesis
that our classification scheme is separating objects with different
natures. 

NGC~6221 provides a local example of a composite object
\citep{Levenson01} with the X-ray flux dominated by the nucleus and  
the visible spectrum dominated by the surrounding starburst. As
\cite{Levenson01} showed, besides X-ray, one can detect the AGN
component in NGC 6221 by the additional broad component of the \oiii\
line in a high S/N nuclear spectrum or with high resolution optical
or NIR imaging. Our objects are much more luminous than such local
examples but otherwise have similar characteristics.

Because the \hb\ emission in X-ray-loud composite galaxies is
dominated by star-forming \hii\
regions, the SFR derived from it are not too far off: they could be 
overestimated by $\sim10\%$.  
As shown in Figure~\ref{fig:hbsfr_x}, the inferred star
formation rates in these galaxies range from a couple to tens of
$\Msun~{\rm yr}^{-1}$ with a median of 10~$\Msun~{\rm yr}^{-1}$, typical of
$z\sim1$ star-forming galaxies \citep{NoeskeWF07} and similar to
the range of SFR found among X-ray selected AGNs at $z\sim0.8$ \citep{Silverman09}. 
The
extinction correction applied is only correct on average but not
accurate for each individual galaxy. We therefore advise against
overinterpreting individual SFR values before better extinction
estimates are made.  

An alternative explanation for the X-ray-loud composite galaxies
might be that  these objects are pure AGNs without star formation, but the X-ray
luminosity is heavily absorbed by a large column density of gas. This
cannot be the case for two reasons. First, it conflicts with
the optical classification. Second, this possibility is not
supported by the X-ray hardness ratio as shown in
Fig.~\ref{fig:hboverx_hr}. The hardness ratio is a very rough indicator
of the X-ray spectral shape, which relates to the level of absorption. 
An unabsorbed X-ray spectrum has a low hardness ratio ($\sim -0.5$). 
Because the opacity is larger for less energetic photons, more absorption 
generally leads to a harder spectrum and a larger hardness ratio. Though 
this correlation is loose and is dependent on redshift \citep{Trouille09}, 
nonetheless, as shown in Fig.~\ref{fig:hboverx_hr}, nearly all of our 
composite galaxies have low hardness ratios, consistent with being unabsorbed.  


The lower panel of Figure~\ref{fig:hboversoftx_hr} shows $L(\hb)/L_X(\soft)$
vs. the hardness ratio. The X-ray-loud composite galaxies
still mainly populate a region different from unambiguous
AGNs, but the separation between the two classes is not as clean
as for the hard band. This is because extinction in the soft band
decreases $L_X(\soft)$ as hardness ratio increases, and the overall
distribution of points shows an overall counterclockwise rotation.


Figure~\ref{fig:hboverx_hr} shows a few X-ray-loud composite galaxies
outside their normal region on the plot.  At least some of them are
likely to be composites of two separate objects rather than a
single composite galaxy.  This is established for one case, DEEP2
object 12016714,\footnote{DEEP2 object number; see http://deep.berkeley.edu/DR1/photo.primer.html .} which has $\hb/L_X(\hard)$ even lower than the typical
value for unambiguous AGNs. {\it HST}/ACS imaging reveals that the single
object in the DEEP2 CFH12K catalog is in fact two galaxies separated
by 1\farcs5. The bluer one is brighter in $R$ and thus was targeted
by DEEP2. However, the X-ray point source is centered on the other,
redder, galaxy.   In total, 14 of our 30
X-ray-loud composite galaxies were imaged by {\it HST}/ACS
\citep{Lotz08}. From visual inspection, two others (DEEP2 12004519,
13049115) out of the 14 are actually close pairs whose
components were not separable in the ground-based images, and we
cannot tell which component contributed either the spectrum or the
X-ray flux. Based on these very rough statistics, we expect 
20\% of all X-ray-loud composite galaxies in our sample could be
unrelated objects that cannot be separated by
the limited resolution (0\farcs6--1\arcsec\ FWHM) of the
ground-based images used for photometry.

\subsection{Nature of XBONGs}


\begin{figure}
\begin{center}
\includegraphics[totalheight=0.35\textheight]{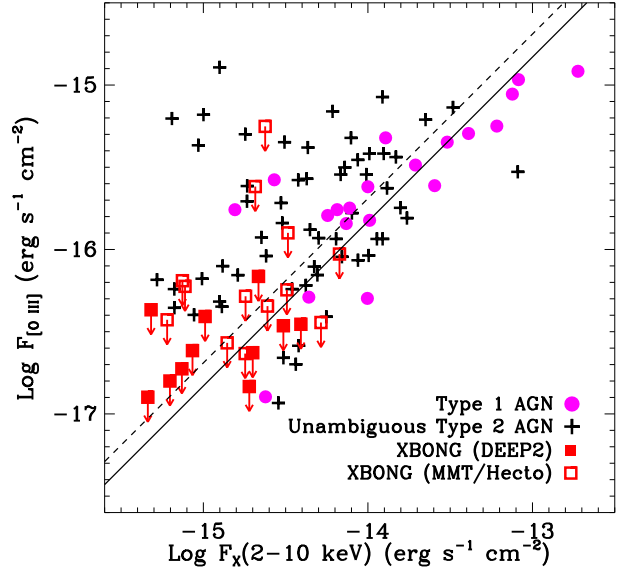}
\caption{\oiii\ line flux (2$\sigma$ upper limit for non-detections)
  vs. X-ray flux at 2-10keV for XBONGs (red squares), Type 1 AGNs
  (magenta circles), and Unambiguous Type 2 AGNs (black crosses). Only
  sources that are detected ($p<0.01$) in the hard band are included in this
  plot.
  The dashed line indicates the median flux ratio found by
  \cite{Heckman05} for Type 1 AGNs, and the solid line indicates the
  median flux ratio for our Type 1 AGN sample. Those XBONGs with DEEP2
  spectra are plotted as solid squares. XBONGs follow the same
  \oiii--X-ray relation as other emission-line AGNs. They do not appear
  to be a physically distinct population.}
\label{fig:o3_x_flux_class3}
\end{center}
\end{figure}

XBONGs are X-ray sources found in
quiescent galaxies for which both \oiii\ and \hb\ are undetected
($<2\sigma$). First, we need to confirm the origin of the X-ray
emission. Besides AGNs, X-ray binaries and hot gas in normal galaxies
can also produce X-ray emission. Most of the galaxies in this category
are red galaxies with early-type morphology.
As shown by \cite{Fabbiano92} and \cite{Hornschemeier05}, the X-ray
luminosity in early types has contributions from both X-ray binaries 
and hot gas, whose total 
luminosity correlates with the stellar mass ($\log L_X \propto 1.8\log
M_*$). We converted the relation given by \cite{Hornschemeier05} to our
bands assuming a thermal Bremsstrahlung spectrum with $T=1$keV and
estimated the expected luminosities for our sources. Only 3 out of
the 32 X-ray sources among quiescent galaxies are both consistent 
with this origin
in hardness ratio and have luminosities (in both soft and hard bands)
within a factor of 3 of the \cite{Hornschemeier05} relation. These
objects could possibly be normal galaxies without active nuclei.
Therefore, we conclude that 29 out of the 32 sources in this category
appear to have their X-ray emission dominated by an AGN.


\cite{YuanN04} argued that XBONGs are powered by a radiatively inefficient
accretion flow resulting in the lack of emission-line regions and
UV/optical bump. Others \citep{Moran02, Trump09} have suggested that
the narrow line emission in these objects is diluted by the host
galaxy, while \cite{Rigby06} argued that heavy extinction in the host
galaxy is responsible for the lack of optical emission lines. However,
none of these analyses has tried to evaluate how much narrow-line
emission is expected given the observed X-ray flux and whether the
non-detections are beyond expectations. For our sample of XBONGs, we
compare their emission line upper limits with their X-ray luminosity
to address this problem.

Here, we only use the term XBONG to refer to galaxies without any
detectable line emission in the DEEP2 or Hectospec spectra. Previous
literature on XBONGs \citep[e.g.][]{Rigby06} has included X-ray
AGNs that are optically classified as star-forming galaxies. As
discussed above, these galaxies do appear to host weak AGNs which are
drowned out by the line emission from star-forming HII regions. Because
the majority of star-forming galaxies are spiral disk galaxies, they will 
show a wide range of axis ratios. The inclusion of these X-ray-loud
composites in the ``optically dull'' AGN sample of \cite{Rigby06} can explain the wide range of axis ratios they found, which were incorrectly used to argue for
host extinction effect. 

In our analysis, we only focus on those sources without any emission
lines, which are sometimes referred to as absorption-dominated,
quiescent, or passive galaxies. 6 of our 29 AGN-dominated XBONGs
actually have \oiiw\ significantly ($>$$2\sigma$) detected. For
consistency, we still count them as XBONGs as if \oii\ were not
covered in the spectra. These objects would likely be classified as
LINERs in a standard BPT diagram as they have very high \oii/\hb\
ratios \citep{Yan06}.

Figure~\ref{fig:o3_x_flux_class3} shows the \oiii\ flux upper
limits vs. hard-X-ray flux distribution for XBONGs in our sample
along with more typical AGNs. The \oiii-to-X-ray ratios of the XBONGs
are consistent with other AGNs. Their \oiii\ upper limits are not low
enough to indicate that they are significantly weaker in their
narrow-line emission relative to their X-ray emission, and they could
simply be the tail of the distribution in \oiii-to-X-ray ratio. 
In fact, many of our XBONGs show weak \oiii\ emission that
is just slightly short of the $2\sigma$ detection threshold. The
median significance of the \oiii\ EW measurement (EW divided by its
uncertainty) among XBONGs is 1.2; 60\% are more than 1$\sigma$ significant.


The XBONGs in our sample have much lower X-ray fluxes than the typical
XBONGs discussed in the literature. All our sources have hard X-ray
flux lower than $6.7\times10^{-15}{\rm erg~s^{-1}~cm^{-2}}$, a factor of four
lower than the prototype XBONG discussed by \cite{Comastri02}, which
has an X-ray flux of $F_{2-10 {\rm keV}}=2.5\times10^{-14}{\rm
  erg~s^{-1}~cm^{-2}}$. A simple explanation for this is that DEEP2 spectra,
with their higher than typical spectral resolution and
signal-to-noise, are able to probe significantly deeper on \oiii\ flux
amid the stellar light and thus reveal optical AGN signatures
for much fainter objects than before. The sample shown here includes
both spectra from the DEEP2 survey and spectra taken in the
MMT/Hectospec follow-up program. The latter data have lower spectral
resolution. If limited to DEEP2-only sources (solid squares in
Fig.~\ref{fig:o3_x_flux_class3}), the XBONGs are even fainter: i.e.,
objects which would be classified as XBONGs in the Hectospec data
yield significant detections if observed by DEIMOS.

Therefore, before we consider any complicated possibilities to explain
XBONGs, we should evaluate the simplest explanation for the
non-detection of \oiii\ in these AGNs: given the observed X-ray flux,
the expected \oiii\ line strength assuming typical AGN flux ratios is
simply beyond our detection capability. Our measurements of \oiii\
upper limits are consistent with this explanation. The \oiii-to-X-ray
flux ratios for our XBONGs are consistent with those of other
narrow-line AGNs and Type 1 AGNs.
They are simply near the tail of the \oiii\ flux distribution at the
corresponding X-ray flux. We do not need to invoke higher than usual
host galaxy extinction to explain them, nor any other physical
mechanism to suppress the narrow-line strength. If these galaxies have
the same \oiii-to-X-ray ratio as Type 1 AGNs, the emission lines would
not have been easily detectable in our spectra. Therefore, we see no
reason to postulate that they are a different type of object, given
the current observations. 

If dilution were the main cause for emission lines in these galaxies to
be undetected, we would expect that XBONGs should have a brighter
rest-frame magnitude in bands bracketing \oiii\ 
than those X-ray sources with similar hard-X-ray
luminosities. 
We investigate this by comparing a sample of our
XBONGs with a sample of unambiguous AGNs matched in hard-X-ray
luminosity. We limit both samples to objects with hard-X-ray (2-10 keV)
luminosity between $10^{41.8} {\rm erg~s^{-1}}$ and $10^{42.8} {\rm erg~s^{-1}}$. 
We can then compare their absolute magnitudes in the continuum
bands bracketing \oiii, which we used to derive the \oiii\ luminosity in
\S\ref{sec:emiline}. It turns out that the two samples have indistinguishable 
distributions in this magnitude. The median \oiii\ sideband magnitudes (AB) for 
the two samples are also very similar: $-20.8$ for the XBONGs and $-20.9$ for 
the unambiguous AGNs.
There is no systematic difference between the two
samples. In only one object---the XBONG with the highest \oiii\ flux
upper limit---is the host galaxy so bright ($M = -23.25$) 
that it is conceivable that dilution could be responsible for the
non-detection. In most cases, dilution by the host galaxy is not
stronger in XBONGs than in other AGN hosts.

\begin{figure}
\begin{center}
\includegraphics[totalheight=0.35\textheight]{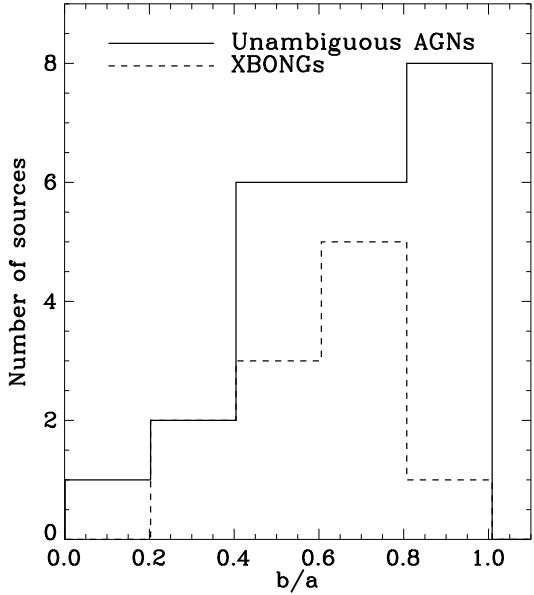}
\caption{Axis ratio distributions for unambiguous AGNs (solid
  histogram) and XBONGs (dashed histogram). Only sources imaged with
  {\it HST}/ACS are included. The two distributions are
  indistinguishable statistically, suggesting that extinction by host
  galaxies is not the primary cause for the nondetection of emission
  lines in XBONGs.}
\label{fig:xbong_ba}
\end{center}
\end{figure}

If extinction in the host galaxies was the main cause for the emission
lines to be undetected, these galaxies would be significantly redder
than other AGN hosts and have smaller axis ratios ($b/a$). We checked
this using the above luminosity-matched comparison sample. The median
$U-B$ color is 1.14 for the XBONGs and 1.07 for the luminosity-matched
unambiguous AGN sample. A Kolmogorov--Smirnov test on the two distributions 
indicates
that the probability of obtaining the observed difference, given the
null hypothesis that the two samples are drawn from the same parent
distribution, is 37\%, meaning the difference is not statistically
significant. To produce such a difference by extinction only requires
an $A_V$ of 0.33 magnitude (assuming $R_V=3.1$), which will only dim
the \oiii\ emission lines by 30\% or 0.15 dex. Therefore, extinction
cannot be the primary reason for the nondetection of emission
lines. Additionally, 11 out of the 29 AGN-dominated XBONGs were imaged
with {\it HST}/ACS. The smallest axis ratio found among them is 0.37\ in the
$F814W$ band. Their axis ratio distribution is indistinguishable from
that of the unambiguous Type 2 AGNs, as shown in Fig.~\ref{fig:xbong_ba}. 

\begin{figure}
\begin{center}
\includegraphics[totalheight=0.35\textheight]{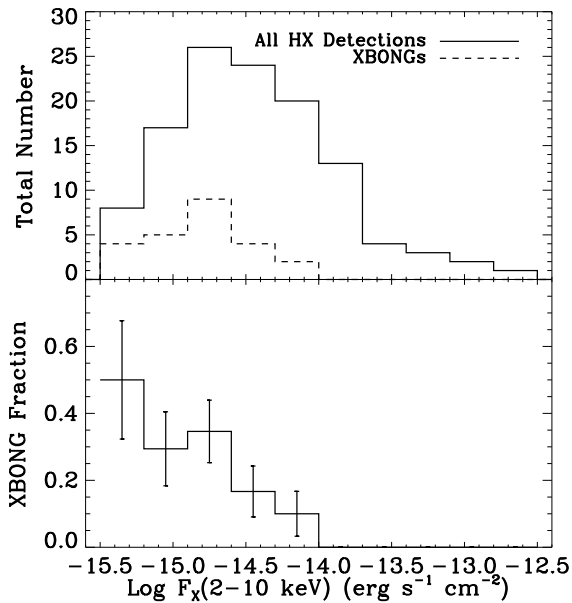}
\caption{Top: the distribution of the 2--10~keV flux for all XBONGs
  (dashed line histogram) detected ($p<0.01$) in the hard X-ray band 
  compared to
  the flux distribution for all hard-X-ray-detected sources (solid
  line histogram). Bottom: the fraction of XBONGs as a function of
  hard X-ray flux. The fraction decreases with increasing flux, which
  suggests that the non-detections are due to the faint X-ray flux of
  the AGNs.}
\label{fig:xbong_frac}
\end{center}
\end{figure}

The fraction of XBONGs in our sample at $0.3<z<0.8$ is
smaller than previously reported in the literature in the same
redshift range ($\sim 30\%$ in \citealt{Trump09a} if limited to $0.3<z<0.8$) and depends strongly on X-ray flux. Among all 146
spectroscopically identified X-ray sources that have both \oiii\ and
\hb\ well covered in the spectra, we have 32 XBONGs, 29 of which are
definitely AGNs. This is $19.9\%\pm3.3\%$. If we limit to DEEP2 spectra
only, which have better quality, the fraction is slightly lower,
$14.7\%\pm3.5\%$ (15 out of 102 sources). Figure~\ref{fig:xbong_frac}
shows the fraction of XBONGs as a function of hard X-ray flux, among
sources that have a measurable hard X-ray flux. The fraction decreases
strongly toward higher X-ray flux. This is consistent with the simple
explanation above that XBONGs are purely the result of observational
limitations rather than comprising a physically distinct class of
AGNs.

We thus find no evidence suggesting that those X-ray sources with no
detected emission-lines must be a separate population from other
emission-line AGNs; neither greater dilution nor higher than usual
host galaxy extinction appears consistent with our observations. To
rule out the null hypothesis that they are the same as other
emission-line AGNs, we need to obtain much higher quality
spectra. Until the time we detect their emission lines and find their
emission-line-to-X-ray ratio is distinctively lower than other AGNs, or
until we push the emission-line upper limits to a correspondingly low
level, there is no reason to classify them separately. Collecting
high-quality spectroscopic data is the best way forward.

\subsection{Incompleteness of the Optical AGN Selection}

The combination of the three classes of objects discussed above
comprises the whole sample of objects that are detected in
$\sim200~{\rm ks}$ $Chandra$ exposures, spectroscopically identified,
brighter than $I_{\rm AB}$ of 22, and have redshifts between
$0.3<z<0.8$. Most of these objects, regardless of their X-ray
luminosity, host an AGN. When classified with optical emission-line
diagnostics, they fall into three classes: emission-line AGNs,
star-forming galaxies, and quiescent galaxies. This reveals the
weakness of optical classification when compared with X-ray
selection. Optical AGN selection not only selects on the bolometric
luminosity of the AGN, it also selects on other properties of the host
galaxy, primarily star formation rate. In the absence of extinction
affecting emission lines and absorption of X-rays, a narrow-line AGN
with a 2--10~keV luminosity of $1.7\times10^{42}{\rm erg~s^{-1}}$ can be
easily drowned out in emission line luminosity by an SFR of $10 {\rm
  M_\odot~yr}^{-1}$, a case in which 97\% of the \hb\ emission comes from
star-forming HII regions, but 97\% of the hard X-ray flux comes from
the AGN.

Most XBONGs should not be counted toward the incompleteness of the 
optical AGN selection, because the intrinsic bolometric luminosity of these
AGNs is simply beyond the detection limit of the optical selection. 
However, the emission line detection limit in the optical spectra is not 
simple to estimate. The detectability depends on many factors: the line flux, 
the stellar continuum flux, the sky background flux, and the complexity of 
the stellar continuum modeling. It also depends on many parameters of the 
observations, such as the exposure time and the seeing at the time of 
observation, which could vary even in the same survey.

%
Optical selection is also sensitive to extinction, which we have not
discussed in detail. For AGN narrow-line regions, it is usually not a
severe concern except in edge-on disk galaxies. 
The median extinction on \oiii\ among typical Type-2 Seyferts is around 
1.0 mag \citep{LaMassa09, Diamond-Stanic09}.  Extinction can be
corrected for when attenuation measurements are available, or one can
exclude edge-on disk galaxies from the analysis. 

Perhaps the most fundamental weakness of optical diagnostics is its 
dependence on high quality spectra, which becomes increasingly expensive
to obtain for fainter galaxies. Many X-ray sources have very faint 
optical counterparts or no counterparts.
In our investigation, we only considered objects with $I_{\rm AB} < 22$ for
completeness and signal-to-noise reasons. In fact, based on photometric 
redshifts obtained from the CFHT Legacy Survey \citep{Ilbert06}, about 
40\% of X-ray sources with an optical counterpart in CFHTLS and with 
$0.3<z_{\rm phot}<0.8$ have $I_{\rm AB} > 22$. Compared 
to our sample, most of them are probably less massive galaxies, which 
have less massive black holes. The AGN demographics of 
these sources could also be different. We leave this for future 
investigations.

\subsection{Optical-only AGNs and the Incompleteness of the X-ray selection}\label{sec:Class 4}

\begin{figure*}
\begin{center}
\includegraphics[totalheight=0.35\textheight]{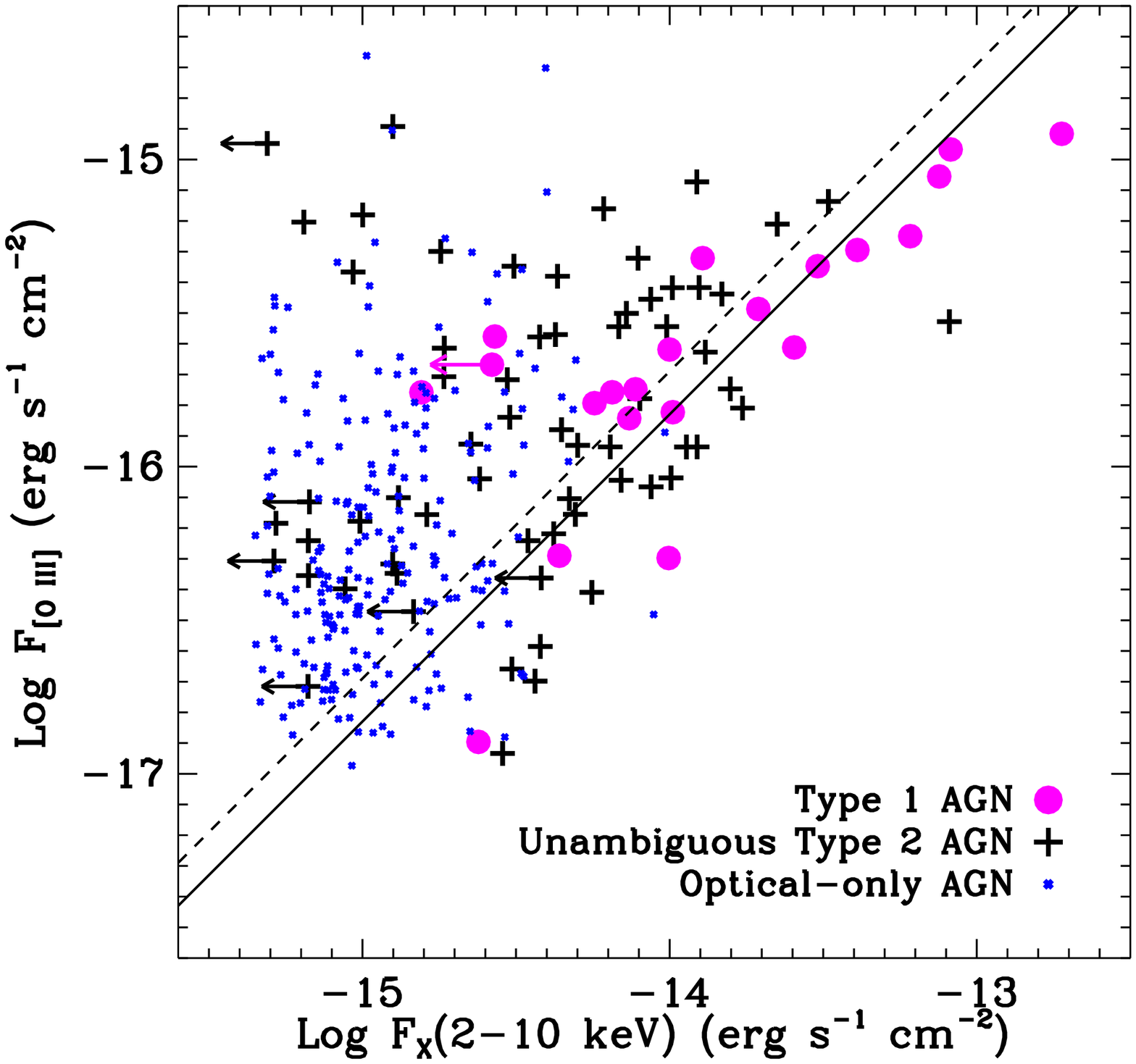}
\includegraphics[totalheight=0.35\textheight]{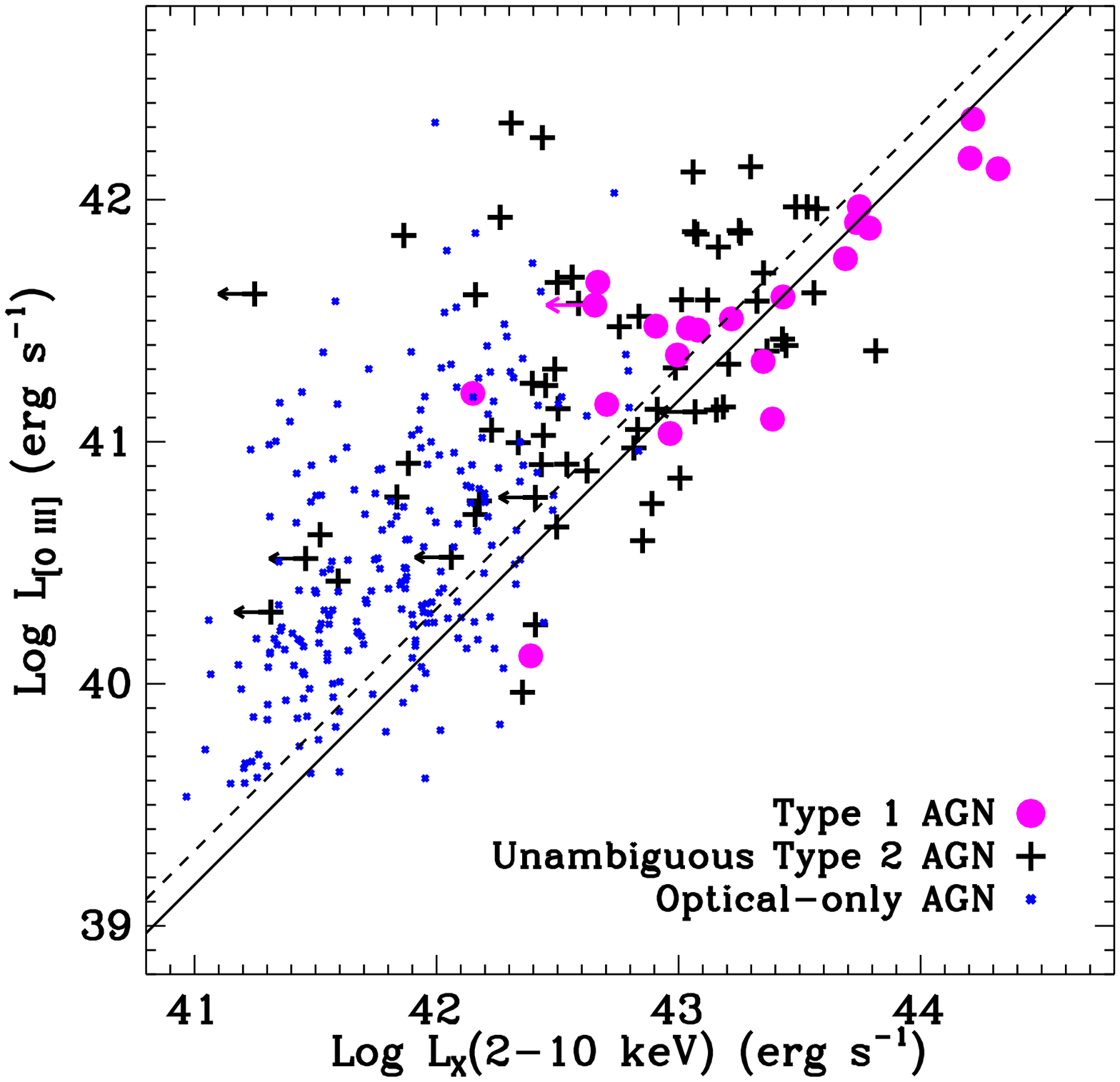}
\caption{Left: \oiii\ line flux vs. X-ray flux in 2--10 keV for Type 1
  AGNs (magenta circles), unambiguous Type 2 AGNs (black crosses), and
  optical-only AGNs (small blue dots); the combination of these
  subsamples forms the optically selected AGN population. Optical-only
  AGNs are undetected in the hard X-ray, thus their upper limits are
  shown without arrows for clarity. The dashed line indicates the
  median flux ratio found by \cite{Heckman05} for Type 1 AGNs. The
  solid line indicates the median flux ratio for our Type 1 sample. As
  seen here, Type 2 AGNs have a broader distribution in \oiii/X-ray
  ratio and mostly have lower hard-X-ray-to-\oiii\ ratios than Type 1
  AGNs. Right: Similar to the left panel but comparing luminosity
  rather than flux.}
\label{fig:o3_x_class1_un}
\end{center}
\end{figure*}

Of the objects which are identified as AGNs from their emission line 
ratios but lacking X-ray
detections, all but one are Type 2 AGNs. Figure~\ref{fig:o3_x_class1_un} 
plots the upper limits for
the 2--10 keV flux and luminosity for these sources along with the
unambiguous AGNs. 
Most of the optical-only AGNs lie to the upper left of the
\cite{Heckman05} relation, i.e., they have much lower
$L_X(\hard)/L_{\oiii}$ ratios. This is consistent with the conclusions
of \cite{Heckman05} based on local AGN samples: optically selected
samples have much lower median $L_X(\hard)/L_{\oiii}$ ratio than X-ray
selected samples and have broader distributions in flux ratio. This
indicates that optically selected samples include more heavily
absorbed sources and possibly Compton-thick sources, which are missed
by X-ray-selection techniques. Therefore, an AGN sample selected based
on a hard-X-ray luminosity threshold in 2--10~keV will not be a complete
bolometric-luminosity-limited sample due to cases of heavy absorption
and Compton-scattering of X-ray photons.

Another potential explanation for the high \oiii-to-X-ray ratio of
these objects is that they have star formation contributing
significantly to the \oiii\ flux but not the X-ray. This cannot be the
case for two reasons. First, these galaxies are classified as AGNs
according to their emission-line ratios indicating that their \oiii\
flux must be dominated by an AGN. Second, star formation would make
these galaxies appear bluer than other AGNs.
The $U-B$ color distribution for optical-only AGNs is statistically
indistinguishable from that of the unambiguous Type 2 AGNs. Therefore,
the high \oiii-to-X-ray ratios of optical-only AGNs cannot be due to
contamination by star formation.

One might worry that these optical-only AGNs are dusty star-forming 
galaxies. For most of them, this cannot be the case. The stellar mass 
distribution of these optical-only AGNs is statistically indistinguishable 
from those AGNs detected in the X-ray (the unambiguous AGNs). On the
other hand, they are much more massive than those star-forming galaxies 
with the same \oiii/\hb\ ratios. The latter has a median stellar mass
of $10^{9.9} M_\odot$, which is only one-tenth of the median mass 
of optical-only AGNs, $10^{11.0} M_\odot$. The difference is much larger
than their respective standard deviations, a factor of 2.8 for the 
star-forming galaxies and a factor of 2.3 for the optical-only AGNs.
The two drastically different stellar mass distributions demonstrate that 
the majority of optical-only AGNs cannot be dusty star-forming galaxies.

Some may also argue that our emission-line selection includes both
Seyferts and LINERs (Low-ionization nuclear emission-line regions),
and some fraction of LINERs could be powered by processes unrelated to
accretion onto SMBHs \citep{Binette94,Sarzi10}. The
recent study by \cite{Sarzi10} using data from the SAURON survey 
showed that ionization processes other than AGN photoionization can
contribute up to 2\AA\ of \oiii\ EW with LINER-like \oiii/\hb\ ratios
in integrated spectra taken with an SDSS fiber aperture.  
Many (35\%, 101 out of 291) of our emission-line-selected AGNs have
\oiii/\hb\ (or lower limits) greater than 3, satisfying the
traditional definition for Seyferts \citep{HoFS97III}. 35\% (67 out of 190) of the remaining objects in
our emission-line AGN sample, which we call LINERs, have \oiii\ EWs
greater than 3\AA, thus definitely having substantial AGN
contribution. In fact, 13\% of those LINERs with \oiii\ EW less than
3\AA\ in our sample are also detected at X-ray wavelengths, suggesting
many of them are indeed AGNs, rather than powered by shocks or old
stellar populations.

\begin{figure}
\begin{center}
\includegraphics[totalheight=0.35\textheight]{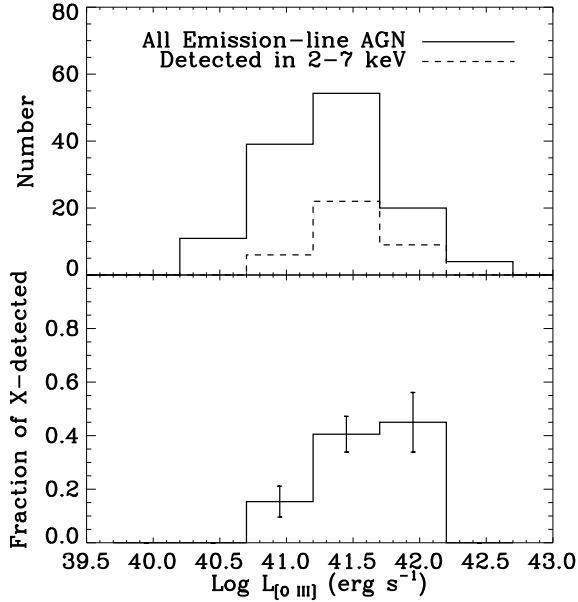}
\caption{Top: the solid histogram shows, in each \oiii\ luminosity
  bin, the sum of the 2--7~keV band detection probabilities of all
  optically selected AGNs (including Type 1s) assuming their X-rays were
  not absorbed. Only DEEP2 objects are included. LINERs (objects
  satisfying the AGN cuts shown in Fig.~\ref{fig:ub_o3hb_xdet_lum} and
  having $\oiii/\hb<3$) with \oiii\ EW less than 3\AA\ are excluded to
  limit contaminations. The dashed histogram shows the number of
  actual 2--7~keV detections ($p_{\rm 2-7~keV}<4\times10^{-6}$) in each \oiii\ 
  luminosity bin. Bottom:
  fraction of 2--7~keV detections among optically-selected AGNs as a
  function of \oiii\ luminosity.}
\label{fig:xfrac_o3lum}
\end{center}
\end{figure}

To evaluate what fraction of genuine AGNs are not detected in the hard X-ray
due to the absorption of the X-ray emission,
we need to take into account the variable sensitivity limit across
each $Chandra$ pointing. Thus, we first estimate how many of the
optically selected AGNs would be detectable in the observed 2--7~keV band 
if they were
not absorbed, and then compare this with the actual number of 2--7~keV 
detections.
We limit this calculation to the DEEP2 optical-AGN sample because the
Hectospec observation gave priorities to X-ray sources in target selection.
We also exclude weak LINER sources with $\oiii/\hb<3$
and ${\rm EW}(\oiii) < 3$\AA\ to limit contamination from
sources not photoionized by an AGN. This is a very conservative AGN
sample. Including both Type 1 and Type 2 optical AGNs, we have a
sample of 140 objects. Our results do not change at all if we strictly limit 
to only Seyferts, i.e., excluding all the LINERs regardless of \oiii\ EW.

Assuming the observed \oiii\ fluxes reflect the intrinsic luminosities of
the AGNs, we estimated the unabsorbed hard-X-ray fluxes for all optical
AGNs using the median hard-X-ray-to-\oiii\ ratio of Type 1 AGNs in our
sample, which is 1.83 dex. 
Given the X-ray exposure map, the background map, and the redshift of
each source, assuming an unabsorbed power-law spectrum with a photon
index of $\Gamma=1.9$, we converted the flux of each source to the
expected source counts in the 2--7~keV band. We then calculated, for each
source, the probability of observing enough counts to qualify it as an
X-ray detection in the hard band, given the background counts at the
position. The sum of these probabilities is the total number of
detectable AGNs if their X-rays were not absorbed at all. Dividing the
number of actual hard-X-ray detections by the sum of the probabilities
yields the X-ray detection fraction, i.e., the fraction of actual
detections out of all potentially X-ray detectable AGNs if the X-rays
were not absorbed. For the 140 objects in the sample defined above,
the sum of their hard band detection probabilities is 128.31. In
reality, only 37 sources (29\%) are detected in the hard band. If we limit
to Seyferts only ($\oiii/\hb>3$), the fraction is the same: out of 91 
Seyferts in our sample, the sum of their potential 2--7~keV detection 
probability is 88.42; while only 26 sources (29\%) are actually detected.

The X-ray detection fraction in bins of \oiii\ luminosity
is plotted in Figure~\ref{fig:xfrac_o3lum}.
The fraction of hard X-ray detection among all potentially-detectable AGNs
decreases toward lower \oiii\ luminosity. At the bright end,
$\sim50\%$ of all AGNs are detected in the 2--7~keV band, which includes
unabsorbed and moderately absorbed AGNs. At the faint end, the
detection rate rolls off because more and more moderately absorbed
AGNs fall below the detection threshold.

This demonstrates the weakness of X-ray selection relative to optical
selection. Depending on the survey depth, X-ray selection can miss 
a substantial population of AGNs
due to absorption and, in some cases, Compton scattering of X-rays by clouds
exterior to the accretion disk but interior to the narrow-line
region. At $L_{\oiii} > 10^{40.5}~{\rm erg~s^{-1}}$, the overall hard X-ray
detection fraction is $29.5\%\pm4.1\%$. Assuming an \oiii\ bolometric
correction of 3500 \citep{Heckman04b}, this corresponds to
$L_{\rm bol}>1.1\times10^{44}~{\rm erg~s^{-1}}$ or intrinsic, unabsorbed
$L_X(\hard)>2.1\times10^{42}~{\rm erg~s^{-1}}$ if the median flux ratio of
Type 1 AGNs is applied. Above this threshold in intrinsic luminosity,
70\% of all potentially detectable AGNs would not be detected (at $p<4\times10^{-6}$) 
in the 2--7~keV band in 200 ks $Chandra$ images due to X-ray absorption and/or 
scattering.  

\subsection{Column density distribution at high-$z$}

Using our emission-line selected AGN sample, we can evaluate whether 
the absorbing column density distribution among high-$z$ AGNs is 
different from that in the local universe. Following most local studies, 
we focus on Seyfert 2 galaxies only.
As shown by \cite{Bassani99}, the column density corresponds closely to 
the HX/\oiii\ ratio. 
By applying the observed hard-X-ray-to-\oiii-ratio distribution of a local 
sample of Seyfert 2s to our high-$z$ sample, we can simulate the expected 
detection fraction of high-$z$ Seyferts if the column density distribution 
among Seyfert 2s does not evolve with redshift.


%

For the local Seyfert 2 sample, we employed the \oiii-selected sample collected 
by \cite{Heckman05}. They provide the observed HX/\oiii\ ratios without any
correction for extinction of \oiii\ or absorption of X-ray, which is ideal
for our purpose. There are 32 Seyfert 2s in this sample, 29 of which
have 2--10~keV X-ray data available. We combined ratios randomly drawn from this
local sample with the observed \oiii\ fluxes of our high-$z$ Seyfert 2s 
to predict their rest-frame 2--10~keV luminosities. With inverse $K$-correction
and conversion from flux to counts (both assuming $\Gamma=1.9$), we predicted the 
observed 2--7~keV counts distribution and the total detection fraction. 
The effect of $\Gamma$ in inverse $K$-correction and the flux-to-counts 
conversion largely cancel out. Assuming the unabsorbed spectral index will 
lead to a slight underestimate\footnote{If true $\Gamma=0$, we will underestimate 
the observed counts at $z=0.6$ by 18\%.} of the observed counts and a lower 
limit on the detection fraction.  
With 5000 simulations, we find the expected 2--7~keV detection 
($p<4\times10^{-6}$) fraction has a mean of 39\% and a dispersion of 5\%. 
In reality, only $25\%\pm5\%$ of our Seyfert 2s are detected in the 2--7~keV 
band, which is $2\sigma$ smaller than expected if the column density 
distribution does not evolve with redshift. This suggests that an average 
Seyfert 2 galaxy between redshift 0.3 and 0.8 has at least the same, or 
marginally higher, column density than the average local Seyfert 2 galaxy. 


We also ran simulations with different detection thresholds to see whether the increased detection fraction of Seyferts will lead to different conclusions. The results are listed in Table.~\ref{tab:simulation}. For the two more relaxed detection thresholds, the differences between the actual detection fraction and the expected detection fraction are smaller and less significant ($\sim1.3\sigma$). Therefore, we conservatively conclude that, at the current statistical significance, the column density distribution among Seyferts at higher-$z$ is similar to that in the local universe, which suggests the fraction of Compton-thick AGNs are also similar to that in the local universe ($\sim50\%$; \citealt{Bassani99},\citealt{Risaliti99}). 

\begin{table}
\begin{center}
\caption{Comparison between actual X-ray detection fraction of Seyfert 2s 
and simulations}
\begin{tabular}{ccc}
\hline \hline
Detection Threshold & Detection Fraction & Simulated Fraction \\ \hline
$p<4\times10^{-6}$  &  $25\%\pm5\%$  & $39\%\pm5\%$ \\
$p<1\times10^{-3}$  &  $37\%\pm5\%$  & $46\%\pm5\%$ \\
$p<1\times10^{-2}$  &  $43\%\pm5\%$  & $52\%\pm5\%$ \\
\hline
\end{tabular}
\label{tab:simulation}
\end{center}
\end{table}




%



\section{Summary and Conclusions}

This paper has developed a new AGN/SF diagnostic diagram using
\oiii/\hb\ ratio and the rest-frame $U-B$ color (in AB system) of the host galaxy. It can be
applied to higher redshifts than more traditional line ratio diagnostics as it
does not require the use of the \nii/\hal\ ratio. 
Using both galaxies at $z\sim0.1$ from the SDSS and galaxies at $0.2<z<0.4$ 
from the DEEP2 survey,
we have demonstrated that this diagnostic technique is highly effective
for galaxies above the Kewley curve in the traditional BPT diagram; but 
less effective for galaxies inbetween the Kauffmann and Kewley demarcations.
All diagrams 
share the same weaknesses, when compared with the X-ray selection.

Applying the new diagram to higher redshifts in the AEGIS survey, we
classified galaxies into AGNs, star-forming galaxies, and quiescent
galaxies. Our sample was selected to have both \oiii\ and \hb\ well
covered in the spectra, which roughly corresponds to the redshift
range $0.3<z<0.8$. We selected only sources with $I_{\rm AB}<22$ that have
secure redshifts, resulting in 3150 objects. Using this sample to compare 
the optical classification to the X-ray data, we have reached the following
conclusions.

\begin{enumerate}

\item $57.5\%\pm4.1\%$ (84 out of 146) of X-ray sources in our sample are
  also emission-line AGNs according to optical selection
  techniques, including both Type 1 and Type 2 objects; $19.2\%\pm3.3\%$
  (28/146) of X-ray sources are classified as star-forming galaxies according to
  our emission-line classification, while $21.9\%\pm3.4\%$ (32/146) are found to
  have neither \oiii\ nor \hb\ detectable, corresponding to X-ray
  bright, optically normal galaxies (XBONGs).

\item For those X-ray sources where the optical emission-line ratios
  indicate star formation, most have X-ray luminosities far exceeding
  the expectations for pure star-forming galaxies. The simplest
  explanation is that they have both an AGN and ongoing star
  formation. 
  Because the \hb-to-X-ray ratio in pure star-forming galaxies is 
  3 orders of magnitude higher than the ratios found in pure AGNs,
  the emission lines in these galaxies are dominated by SF, and the X-ray
  emission is mainly powered by an AGN. 

  Combining this emission-line-to-X-ray ratio with the hardness ratio
  allows us to exclude the possibility of heavily obscured AGNs and to
  disentangle the contributions from AGNs and star formation.

\item In our sample, 21.9\% of X-ray-detected galaxies are found to
  lack both \oiii\ and \hb\ emission lines, which would cause them to
  be classified as XBONGs. All but 3 of them have X-ray luminosities
  exceeding the expectations for normal early-type galaxies,
  indicating the presence of AGNs. 
  These sources have \oiii\ upper limits consistent with the
  expectation from the X-ray luminosity for typical AGNs, i.e., they
  are not distinctively lower in their \oiii-to-X-ray ratios. There is
  no reason to assume that XBONGs are a physically different
  population from other X-ray AGNs. Neither 
  host galaxy dilution nor unusual extinction is primarily
  responsible for the non-detection of line
  emission in most of the XBONGs. 


\item 
  Our new emission line ratio diagnostics identifies 291 AGNs in our sample, 
  of which 22\% are 
  also detected in the X-ray sample. Absorption of the X-rays by gas near the 
  SMBH is necessary to account for most of the non-detections. 
  Taking into account the variable
  sensitivity across $Chandra$ pointings, we estimated the X-ray
  detection fraction as a function of the observed \oiii\
  luminosity. At $L_{\rm bol} > 10^{44} {\rm erg~s^{-1}}$, about 2/3 of the
  emission-line AGNs with $0.3<z<0.8$ and $I_{\rm AB}<22$ will not be detected
  in the 2--7~keV band in our $\sim200~{\rm ks}$ $Chandra$ images due 
  to absorption and/or scattering of the X-rays.

\item 
  If the column density distribution of Seyfert 2 galaxies at high $z$ were the 
  same as in the local universe, we would expect a slightly higher fraction 
  of our Seyfert 2s to be detected in the 2--7~keV band than observed. 
  This suggests that
  Seyfert 2 galaxies at $0.3<z<0.8$ have the same or marginally higher 
  average column density than local Seyfert 2s. Thus, we expect the
  Compton-thick fractions at both redshifts to be similar as well.

\end{enumerate}

Neither optical classification nor X-ray selection yields a complete
AGN sample; in fact, both are far from that goal. In the X-ray, heavy
absorption by gas in close proximity to the AGN can prevent the
detection of a substantial population of AGNs. The optical selection is
less affected by obscuring material as the narrow emission lines arise
from much larger scales. However, emission lines can easily be drowned
out by star formation. Additionally, the detection of line emission
requires high quality spectra to subtract the host galaxy stellar
light. The combination of the two methods gives a more complete
sample. However, heavily X-ray-absorbed AGNs that reside in
star-forming galaxies will still be missed. Infrared observations
could be the solution to finding AGNs in these cases
\citep{Lacy04,Stern05, Park10}.


\bigskip
\acknowledgements
This paper is based on observations taken at the W. M. Keck Observatory 
which is operated 
jointly by the University of California and the California Institute of 
Technology, and the MMT Observatory , a joint facility of the Smithsonian 
Institution and the University of Arizona. 

We thank the anonymous referee who helped us to improve
the paper significantly. We thank Greg Wirth and all of 
the Keck Observatory
staff for their help in the acquisition of the DEEP2 Keck/DEIMOS data. We
also wish to recognize and acknowledge the highly significant cultural
role and reverence that the summit of Mauna Kea has always had within
the indigenous Hawaiian community. It is a privilege to be given the
opportunity to conduct observations from this mountain.

This study makes use of data from AEGIS, a multiwavelength sky survey
conducted with the $Chandra$, GALEX, Hubble, Keck, CFHT, MMT, Subaru,
Palomar, {\it Spitzer}, VLA, and other telescopes and supported in part by
the NSF, NASA, and the STFC.
The AEGIS website is http://aegis.ucolick.org . The DEEP2 website is
http://deep.berkeley.edu/ .

The project was supported in part by the NSF Grants AST00-71198, 
AST00-71048, AST05-07483, AST05-07428, AST08-07630, AST08-08133, and 
NASA Chandra grants G05-6141A and GO8-9129A. 
This research made use of the NASA Astrophysics Data System,
and employed open-source software written and maintained by David Schlegel,
Douglas Finkbeiner, and others. 

Funding for the Sloan Digital Sky Survey (SDSS) has been provided by
the Alfred P. Sloan Foundation, the Participating Institutions, the
National Aeronautics and Space Administration, the National Science
Foundation, the U.S. Department of Energy, the Japanese
Monbukagakusho, and the Max Planck Society. The SDSS Web site is
http://www.sdss.org/.
The SDSS is managed by the Astrophysical Research Consortium (ARC) for
the Participating Institutions. The Participating Institutions are The
University of Chicago, Fermilab, the Institute for Advanced Study, the
Japan Participation Group, The Johns Hopkins University, Los Alamos
National Laboratory, the Max-Planck-Institute for Astronomy (MPIA),
the Max-Planck-Institute for Astrophysics (MPA), New Mexico State
University, the University of Pittsburgh, Princeton University, the
United States Naval Observatory, and the University of Washington.

\facility{
{\it Facilities:} \facility{Keck:II (DEIMOS)}, \facility{MMT (Hectospec)}, \facility{CXO (ACIS)}, \facility{CFHT (CFH12K)}, \facility{HST (ACS)}
}

\bibliographystyle{apj}
\bibliography{apj-jour,astro_refs}

\include{objproptable}

\end{document}

%% file: objproptable.tex
\clearpage
\begin{landscape}
\scriptsize
\begin{center}
\begin{longtable}{ccccccccccccc}
\caption{Optical properties of all X-ray sources and optically classified AGNs in our sample} \label{tab:xobj_prop} \\
\hline \hline \\[0.5ex]
\multicolumn{1}{c}{\textbf{EGSAGN}} &
\multicolumn{1}{c}{\textbf{LNG2009}} &
\multicolumn{1}{c}{\textbf{DEEP2}} &
\multicolumn{1}{c}{\textbf{RA}} &
\multicolumn{1}{c}{\textbf{Dec}} &
\multicolumn{1}{c}{\textbf{$z$}} &
\multicolumn{1}{c}{\textbf{Mag $I$}} &
\multicolumn{1}{c}{\textbf{$U-B$}} &
\multicolumn{1}{c}{$\log F_{\rm [O~III]}$} &
\multicolumn{1}{c}{$\log F_{\hb}$} &
\multicolumn{1}{c}{$\log (\oiii/\hb)$} &
\multicolumn{1}{c}{\textbf{Survey}} &
\multicolumn{1}{c}{\textbf{Classification}} \\ 
\multicolumn{1}{c}{No} &
\multicolumn{1}{c}{\textbf{ID} (1)} &
\multicolumn{1}{c}{\textbf{ObjNo} (2)} &
\multicolumn{2}{c}{\textbf{(J2000)}} &
\multicolumn{1}{c}{} &
\multicolumn{1}{c}{} &
\multicolumn{1}{c}{} &
\multicolumn{1}{c}{(${\rm erg/s/cm^2}$)} &
\multicolumn{1}{c}{(${\rm erg/s/cm^2}$)} &
\multicolumn{1}{c}{} &
\multicolumn{1}{c}{(3)} &
\multicolumn{1}{c}{(4)} \\[0.5ex] \hline
\\[0.5ex]
\endfirsthead

\hline \hline \\[0.5ex]
\multicolumn{1}{c}{\textbf{EGSAGN}} &
\multicolumn{1}{c}{\textbf{LNG2009}} &
\multicolumn{1}{c}{\textbf{DEEP2}} &
\multicolumn{1}{c}{\textbf{RA}} &
\multicolumn{1}{c}{\textbf{Dec}} &
\multicolumn{1}{c}{\textbf{$z$}} &
\multicolumn{1}{c}{\textbf{Mag $I$}} &
\multicolumn{1}{c}{\textbf{$U-B$}} &
\multicolumn{1}{c}{$\log F_{\rm [O~III]}$} &
\multicolumn{1}{c}{$\log F_{\hb}$} &
\multicolumn{1}{c}{$\log (\oiii/\hb)$} &
\multicolumn{1}{c}{\textbf{Survey}} &
\multicolumn{1}{c}{\textbf{Classification}} \\ 
\multicolumn{1}{c}{\textbf{No}} &
\multicolumn{1}{c}{\textbf{ID} (1)} &
\multicolumn{1}{c}{\textbf{ObjNo (2)}} &
\multicolumn{2}{c}{\textbf{(J2000)}} &
\multicolumn{1}{c}{} &
\multicolumn{1}{c}{} &
\multicolumn{1}{c}{} &
\multicolumn{1}{c}{(${\rm erg/s/cm^2}$)} &
\multicolumn{1}{c}{(${\rm erg/s/cm^2}$)} &
\multicolumn{1}{c}{} &
\multicolumn{1}{c}{(3)} &
\multicolumn{1}{c}{(4)} \\[0.5ex] \hline
\\[0.5ex]
\endhead

\\[0.5ex] \hline 
\\[0.5ex]
\multicolumn{3}{l}{{Continued on Next Page\ldots}} \\
\endfoot

\\[0.5ex] \hline\hline
\endlastfoot

   1 &egs\_0079 &11007255 &213.82423 &51.97931 &  0.6505 &20.51 & 1.17 &$ -16.41\pm0.14$ &$< -16.75$ &$> 0.34$ &  1 &AGN-2 \\
   2 &          &11007325 &213.81808 &51.98800 &  0.5308 &20.58 & 1.23 &$ -16.44\pm0.20$ &$< -16.63$ &$> 0.15$ &  1 &AGN-2 \\
   3 &          &11007338 &213.86714 &51.97939 &  0.4278 &19.97 & 1.20 &$ -16.40\pm0.17$ &$< -16.62$ &$> 0.21$ &  1 &AGN-2 \\
   4 &          &11007806 &213.79913 &52.01271 &  0.5152 &20.57 & 1.26 &$ -16.52\pm0.19$ &$< -16.77$ &$> 0.23$ &  1 &AGN-2 \\
   5 &          &11013115 &213.95496 &52.05116 &  0.6227 &21.08 & 0.80 &$ -16.24\pm0.07$ &$ -16.64\pm0.21$ &$  0.45$ &  1 &AGN-2 \\
   6 &egs\_0177 &11013281 &213.97916 &52.05984 &  0.4477 &19.49 & 1.06 &$< -16.25$ &$< -16.61$ &... &  2 &XBONG \\
   7 &egs\_0107 &11013724 &213.88416 &52.06547 &  0.6498 &21.04 & 0.89 &$< -16.53$ &$< -16.58$ &... &  1 &XBONG \\
   8 &          &11013883 &213.81531 &52.06502 &  0.6934 &20.65 & 1.04 &$ -15.75\pm0.03$ &$< -16.84$ &$> 1.19$ &  1 &AGN-2 \\
\end{longtable}
\end{center}
\begin{enumerate}[(1)]
\item ID in \citep{Laird09}. X-ray undetected sources have no ID.
\item DEEP2 object number; see http://deep.berkeley.edu/DR1/photo.primer.html  .
\item Source of the redshift, \oiii, and \hb\ measurements: 1 -- DEEP2, 2 -- MMT/Hectospec follow-up.
\item Classsifications: AGN-1 = Type 1 AGN; AGN-2 = optically selected Type 2 AGN, including both X-ray detected and undetected; SF+AGN = X-ray loud, composite galaxies; XBONG = X-ray Bright, Optically Normal Galaxies; Gal = X-ray detected normal galaxies; Ambiguous = X-ray sources with ambiguous optical classifications; Blended = known cases of blended objects for which the optical spectrum and the X-rays are from different objects.
\item DEEP2 Object 11027275 is a B-band drop out; its U-B color is not available.
\end{enumerate}
\clearpage
\end{landscape}